\documentclass[12pt,nofootinbib,floatfix]{revtex4}
\pdfoutput=1

\usepackage{empheq} 
\usepackage{mathrsfs}
\usepackage{verbatim}
\usepackage{amssymb}
\usepackage{graphicx}
\usepackage{epsfig}
\usepackage{color}
\usepackage{rotating}
\usepackage{amsbsy}

\usepackage[mathscr]{euscript}
\usepackage{amsmath}
\usepackage{textcomp}
\usepackage{ulem}

\def\leq{\,\raise 0.4ex\hbox{$<$}\kern -0.8em\lower 0.62ex\hbox{$-$}\,}
\def\geq{\,\raise 0.4ex\hbox{$>$}\kern -0.8em\lower 0.62ex\hbox{$-$}\,}
\def\pm{\,\raise 0.4ex\hbox{$+$}\kern -0.8em\lower 0.62ex\hbox{$-$}\,}
\def\lsim{\,\raise 0.4ex\hbox{$<$}\kern -0.8em\lower 0.62ex\hbox{$\sim$}\,}
\def\gsim{\,\raise 0.4ex\hbox{$>$}\kern -0.8em\lower 0.62ex\hbox{$\sim$}\,}

\def\bhPhi{{\bf {\hat \Phi}}}
\def\bhPsi{{\bf {\hat \Psi}}}

\def\bv{{\bf v}}
\def\br{{\bf r}}
\def\bp{{\bf p}}
\def\bn{{\bf n}}
\def\bl{{\bf L}}
\def\bL{{\bf L}}
\def\bhl{{\bf {\hat L}}}
\def\bs{{\bf S_{eff}}}
\def\bsone{{\bf S_1}}
\def\bstwo{{\bf S_2}}

\def\bstot{{\bf S}}
\def\bri{{\bf \xi}_{ADM}}

\def\bsL{{\bf {\mathscr S}}}
\def\bsoneL{{\bf \mathscr S_1}}
\def\bstwoL{{\bf \mathscr S_2}}

\def\bj{{\bf J}}
\def\bJ{{\bf J}}
\def\bhj{{\bf {\hat J}}}
\def\bn{{\bf {\hat n}}}

\def\bx{{\bf x}}

\def\blnt{{\bf \tilde {\bf L}}_{N}}
\def\bxi{{\bf \xi}}

\def\avec{\bf {\it a}}

\def\be{\begin{equation}}
\def\ee{\end{equation}}
\def\ba{\begin{eqnarray}}
\def\ea{\end{eqnarray}}

\newcommand{\half}[0]{\frac{1}{2}}

\begin{document}
\bibliographystyle{apsrev}

\title{Inspiral of Generic Black Hole Binaries:\par Spin, Precession, and Eccentricity}

\author{Janna Levin${}^{1,2}$, Sean T. McWilliams${}^{1,2,3}$, and Hugo Contreras${}^4$}
\affiliation{${}^{1}$Department of Physics and Astronomy, Barnard
College of Columbia University, 3009 Broadway, New York, NY 10027 }
\affiliation{${}^{2}$Institute for Strings, Cosmology and Astroparticle
  Physics (ISCAP), Columbia University, New York, NY 10027}
\affiliation{${}^{3}$Department of Physics, Princeton University, Princeton, NJ 08544}
\affiliation{${}^{4}$Department of Physics, Columbia University, New York, NY 10027}

\email{janna@astro.columbia.edu}

\begin{abstract}
Given the absence of observations of black hole binaries, 
it is critical that the full range of accessible parameter space be explored in anticipation
of future observation with gravitational wave detectors.
To this end, we compile the Hamiltonian equations of motion describing the conservative dynamics of the most general black
hole binaries and incorporate an effective treatment of dissipation
through gravitational radiation, as computed by Will and collaborators.
We evolve these 
equations for systems with orbital eccentricity and precessing spins.
We find that, while
spin-spin coupling corrections can destroy constant radius orbits in
principle, the effect is so small that orbits will reliably tend to quasi-spherical orbits
as angular momentum and energy are lost to gravitational radiation. 
Still, binaries that are initially highly eccentric may retain eccentricity as they pass into the detectable bandwidth
of ground-based gravitational wave detectors.
We also show that a useful set of
natural frequencies for an orbit demonstrating both spin
precession and periastron precession is comprised of (1) the frequency of angular
motion in the orbital plane, (2) the frequency of the plane precession,
and (3) the frequency of radial oscillations. These three natural harmonics shape the
observed waveform.
\end{abstract}
\maketitle

\section{Introduction}

Motivated by future gravitational-wave observatories, a campaign
to track the inspiral of the most generic black hole binaries
has long been underway
\cite{{schaefer1985},{damour1981},{damour1988},{jaranowski1998},{damourpn2001},{damourpn2000:2},PhysRevD.52.6882,Mora:2003wt,Pati:2002ux,Will:2005sn,Wang:2007ntb}. 
The promise of
gravitational-wave astronomy lies in our ability to observe and test
the full range of astrophysical phenomena, including spinning, precessing,
eccentric pairs of unequal mass. To this end, 
Will and collaborators have published a series of 
computations of
the equations governing black hole binary motion in the Post-Newtonian
(PN) expansion to 3.5PN order, including spin corrections
\cite{Mora:2003wt,Pati:2002ux,Will:2005sn,Wang:2007ntb}. We compile
those results in an appendix  to provide a resource for probing and testing the
PN dynamics. We then convert the dissipative terms into Hamiltonian
variables and suggest a modification of the Hamiltonian equations of
motion that incorporates the effects of radiation reaction. 
The modified Hamiltonian formulation and the Lagrangian formulation admit equivalent
descriptions, though 
for the purposes of 
investigating the natural harmonics of these systems we will exploit
the ease of interpretation offered by the Hamiltonian formulation.

Binary stars that evolve to a pair of black holes can show evidence of
eccentricity and spin precession in waveforms detectable by LISA \cite{sesana2010}.
Although long-lived pairs will likely shed eccentricity
by the time they enter the LIGO bandwidth, the entire
orbital plane continues to precess along with the spins. Also, black hole
pairs formed by tidal capture in globular clusters or galactic nuclei
may retain significant eccentricity as their signals pass through the band of current and future ground-based
gravitational-wave observatories 
\cite{wen2003,O'Leary:2008xt}. The equations of motion of
Refs.\ \cite{Mora:2003wt,Pati:2002ux,Will:2005sn,Wang:2007ntb} allow flexibility in
handling the gravitational radiation emitted by any realistic black
hole pair, prior to entering the strong-field.
 
There are four questions we can address immediately with this
compilation of the inspiral equations: (1) 
Do spinning pairs tend to quasi-spherical orbits?
(2) What 
features are generically introduced into waveforms through periastron
precession and spin precession?
(3) How much
energy is lost
during each burst near periastron passage?
(4) How much of the orbit and the waveform for eccentric, precessing orbits
is well-described by the PN expansion?

The first question
(Do spinning pairs tend to quasi-spherical orbits?) must be asked since the purely circular orbits are
destroyed by spin-spin (SS) couplings.
When spin-orbit (SO) coupling is incorporated, the entire orbital-plane precesses and the
quasi-circular orbits are replaced by quasi-spherical orbits --
trajectories that lie on the surface of a sphere whose radius shrinks
only due to dissipation. However, spin-spin couplings actually destroy even
these so that there are no constant radius orbits, even if we were to
artificially turn off dissipation by turning off the half-order terms
in the expansion. In other
words, if black holes spin, {\it there may not be any quasi-spherical} orbits
and all orbits could retain eccentricity at all stages of their inspiral. 
We can ask how significant the effect is. Since spin-spin is a
subdominant effect, we find that the eccentric
behavior is small and that orbits can appear to be very nearly
quasi-spherical.

The second question
(What
features are generically introduced into waveforms through periastron
precession and spin precession?) is
significant for designing optimal detection algorithms and estimating source parameters.
For systems possessing spin, modulation due to orbital plane precession will leave an imprint on the observed waveform.
In the case of significant eccentricity, the waveforms are
modulated by the radial oscillation of the orbit. Characteristically the amplitude is modulated by
eccentricity, as is the polarization due to the precession of the
periastron and the orbital plane. The Fourier
transform of the waveform will reflect these precessions by
reflecting the natural frequencies of the orbit, which modulate the frequency evolution from loss of orbital energy
that is present in all black hole binaries. We demonstrate
these features for an example black hole binary that possesses all three characteristic frequencies, for which we also
address the third question posed (How much
energy is lost during each burst near periastron passage?).

The fourth question (How much of the orbit and waveform for eccentric, precessing orbits
is well described by the PN expansion?) has not been addressed for generic orbits.  In the
limit of quasi-spherical orbits, the dynamics and waveform are usually taken to be accurate
until the system reaches the innermost stable circular orbit (ISCO), beyond which even the
conservative dynamics cannot be treated adiabatically.  However, the ISCO is formally a characteristic
of a test-particle orbit in Schwarzschild or Kerr spacetime, and is not well defined for binary spacetimes.  As we will
show, the PN sequence actually diverges well outside the ISCO not only for eccentric precessing systems,
but for quasi-spherical systems as well.
We find that the breakdown happens at radial separations $r \sim 10M$,
which is well outside the Schwarzschild ISCO ($6M$), in contrast to the
conventional wisdom of using $6M$ as a reference point for truncating PN approximations.

Due to this limitation of the approximation, the PN equations of
motion cannot be used to probe the most extreme form of precession
manifest as
zoom-whirl behavior -- elliptical zooms out to apastron followed by multiple nearly circular
whirls around periastron \cite{Glampedakis:2005hs}.  Zoom-whirl orbits are most
prevalent when periastron drops into the strong-field regime. As
shown in Ref.\ \cite{Levin:2008mq}, complete whirls occur when periastron falls between
the IBCO (innermost bound circular orbit) and the ISCO.
To be clear, zoom-whirl orbits exist and have been observed in
numerical relativity simulations \cite{pretorius2007,Healy:2009zm}; they
are simply beyond the trusted regime for the PN approximation.

To lay the foundation, we begin with a discussion of conservative
black hole dynamics in
\S \ref{sec:dynamics}.
In \S \ref{sec:gallery}, we discuss the utility of a Hamiltonian
formulation of the equations of motion for simplifying interpretation of the dynamics of
the full inspiral trajectories
and resulting gravitational waveforms for
spinning black hole pairs on eccentric, precessing orbits.
In \S \ref{sec:valid}, we explore the limits of our, and indeed any, PN formulation,
due to the intrinsic non-perturbative nature of the dynamics at the end of the inspiral.
In \S
\ref{sec:summary}, we summarize the results and the answers to the
four questions we posed.
For completeness and ease of reference, in \S \ref{sec:lag} the PN corrections to the
equations of motion including the dissipation due to gravitational
radiation are compiled from Refs.\ \cite{Mora:2003wt,Pati:2002ux,Will:2005sn,Wang:2007ntb}
 as derived by Will and
collaborators to 3.5PN order with SO
and SS coupling. Finally, in \S \ref{sec:ham}, we convert the PN corrections
originally computed in the Lagrangian formulation into Hamiltonian
coordinates, which is the approach we apply throughout the text.

\section{Dynamics Without Dissipation}
\label{sec:dynamics}

For this section, radiation reaction has artificially been
turned off so the pair shows no evidence of dissipation. The advantage
of doing so is that we can 
clearly see intrinsic features of the dynamics that
dissipation can obscure. In the following section, dissipation is included
while the insights of this section will continue to guide our perspective.

\begin{figure}
\includegraphics[trim = 0mm 10mm 0mm 30mm, width=0.35\textwidth]{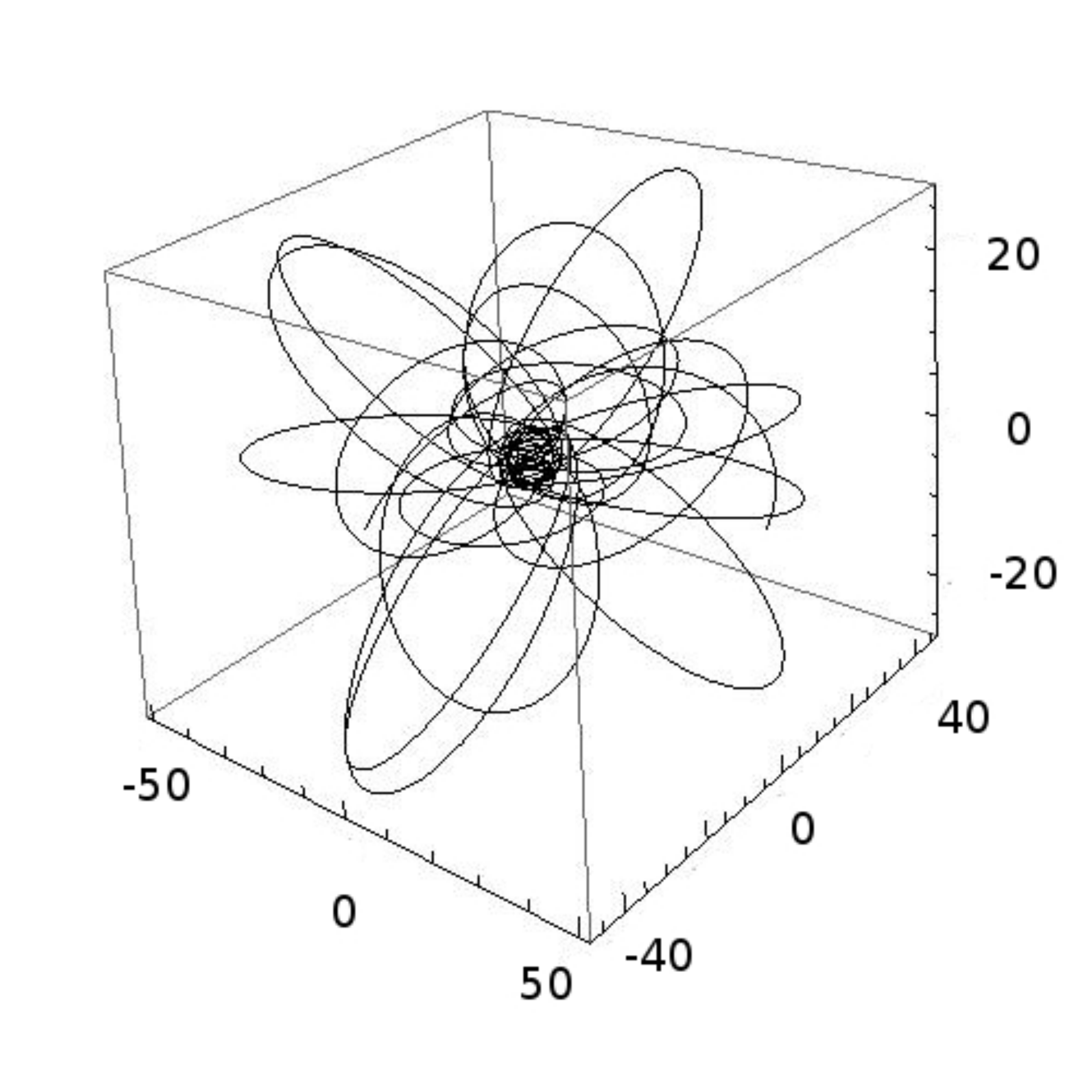}
\centering
\includegraphics[trim = 0mm 0mm 0mm 30mm, width=0.28\textwidth]{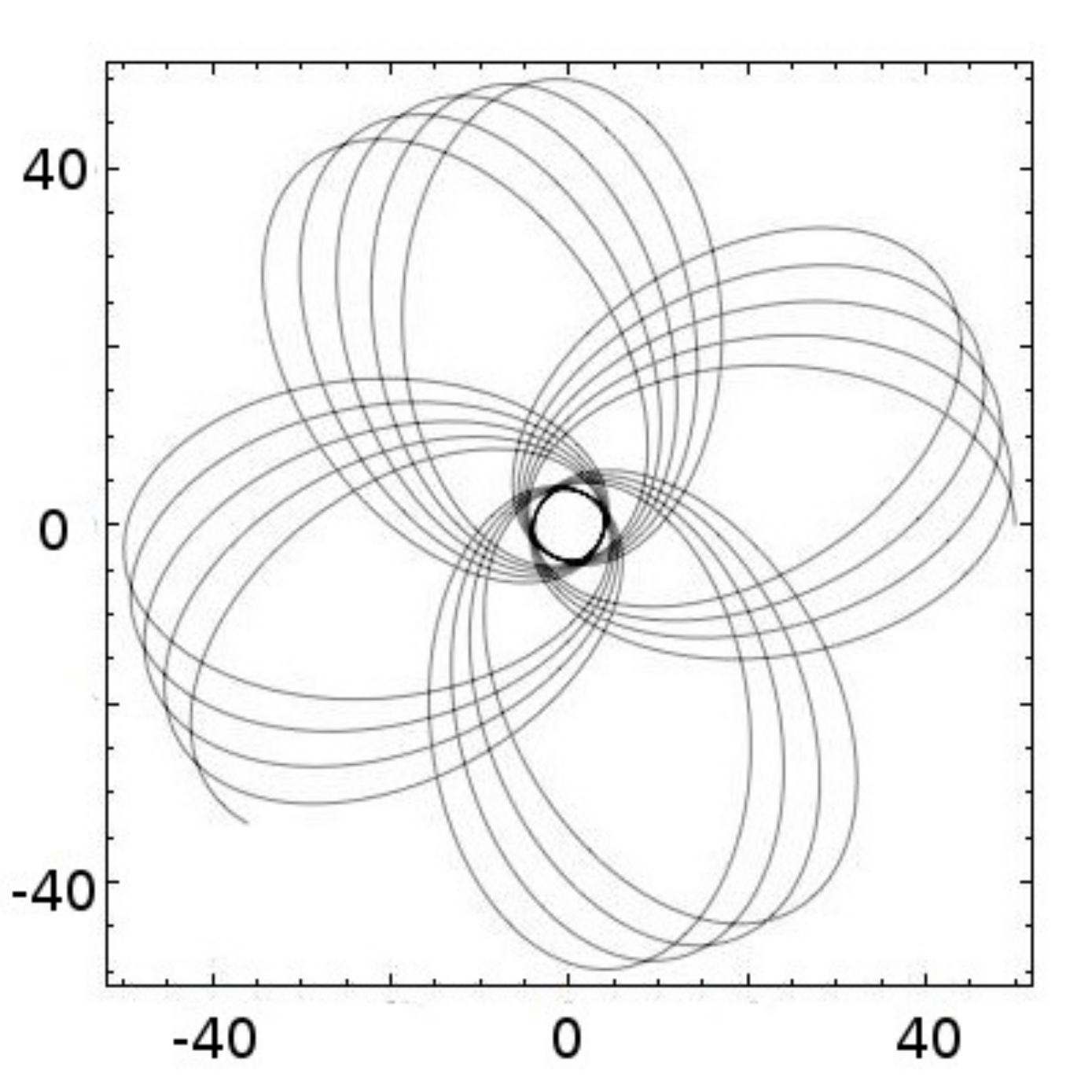}
\centering 
\includegraphics[trim = 0mm 5mm 0mm 0mm, width=0.3\textwidth]{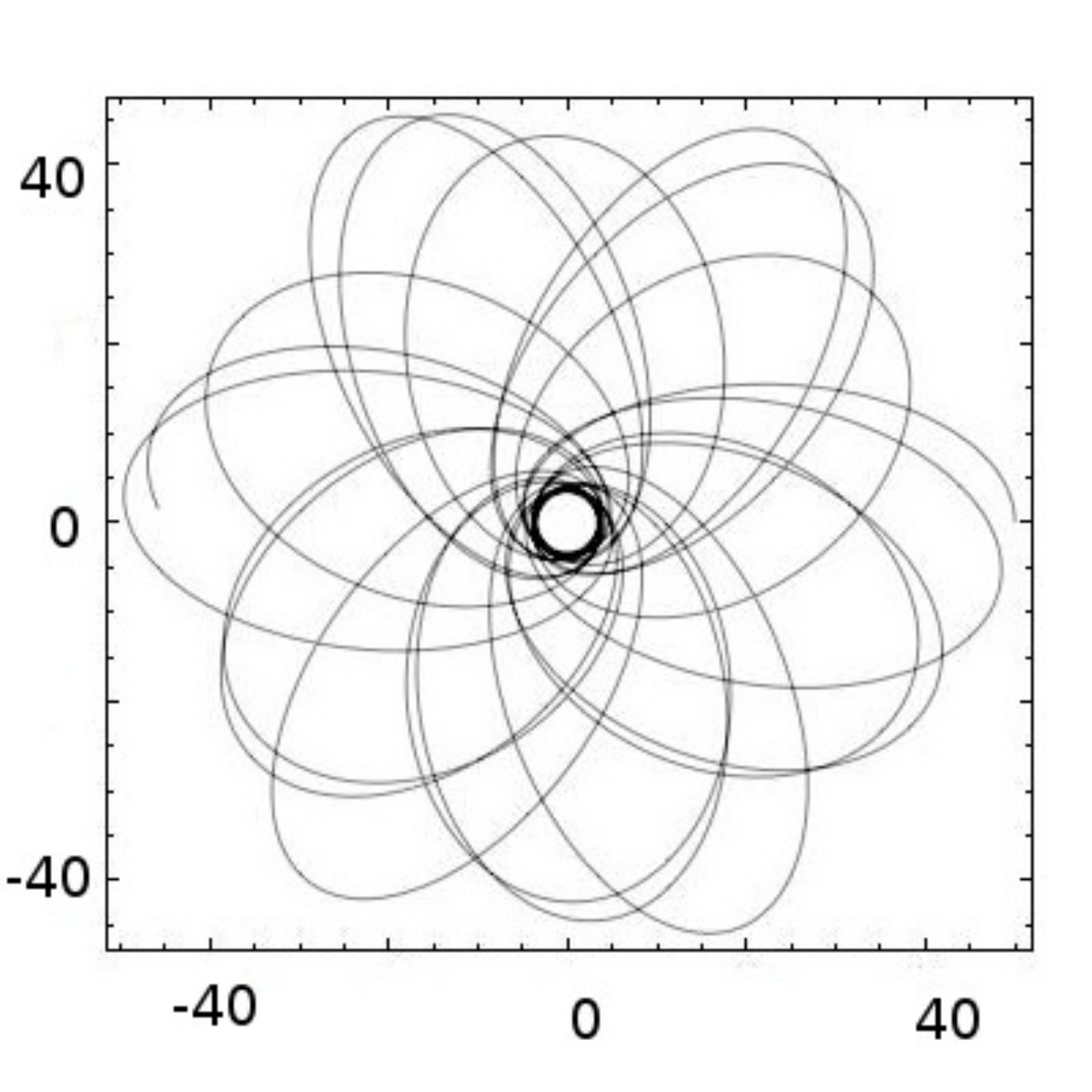}
\centering
\hfill
  \caption{A black hole pair with $m_2/m_1=1/4$. The heavier of the
    two black holes is spinning at $0.9$ maximal. The lighter is not
    spinning. 
Left: The orbit. 
Middle: 
The orbit in the orbital plane
    showing the constancy of the periastron and apastron as well as
    isolating the periastron precession. In the orbital plane, the
    pair precesses around a four-leaf clover. 
Right: A projection of the orbit onto
    the equatorial plane, showing that the innate symmetry of the orbit is obscured. 
  \label{Fig:Conservative}
}
\end{figure}

\begin{figure}
 \includegraphics[trim = 50mm 50mm 14mm 50mm, width=0.49\textwidth]{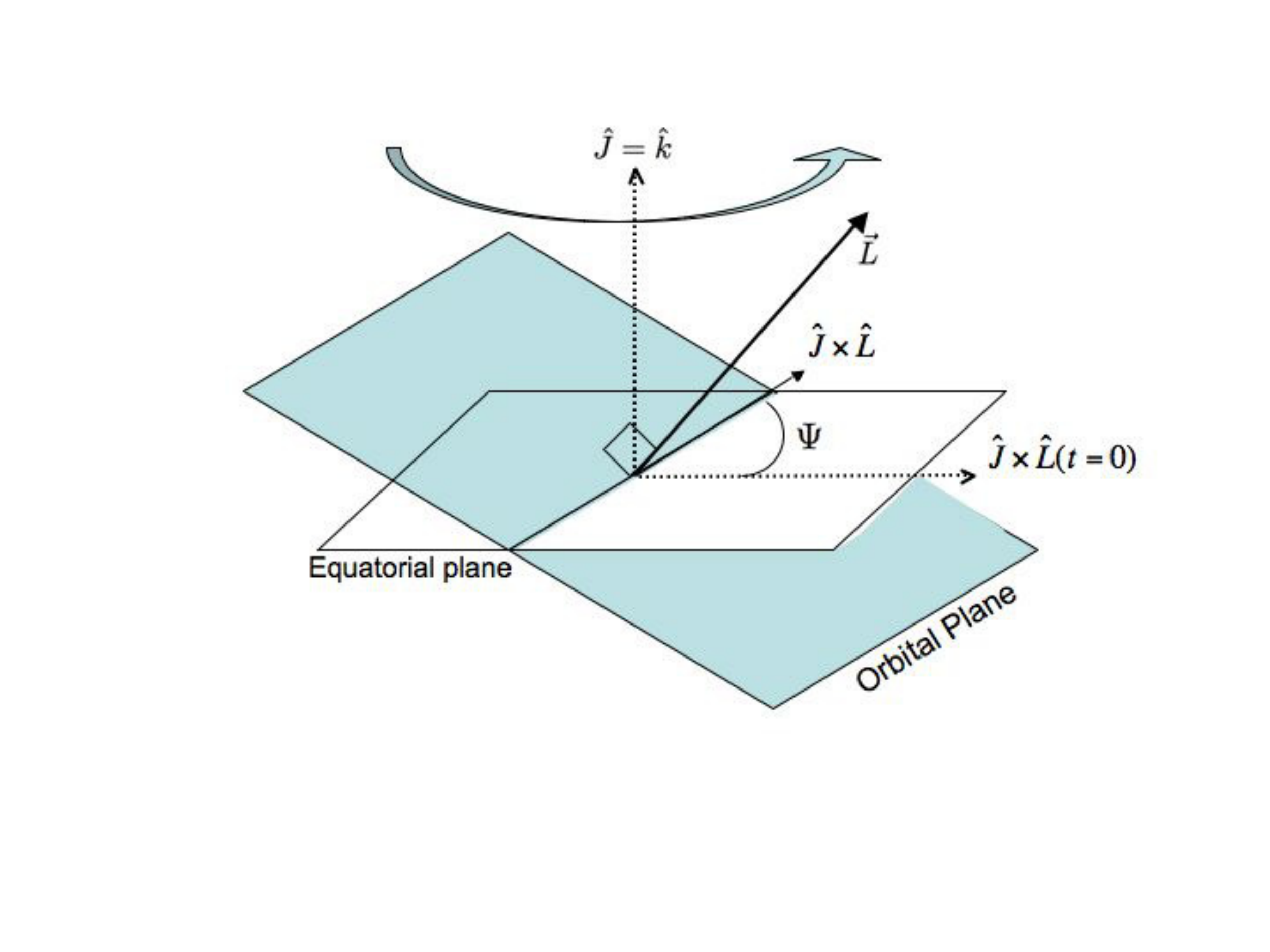}
 \includegraphics[trim = 14mm 50mm 50mm 50mm, width=0.5\textwidth]{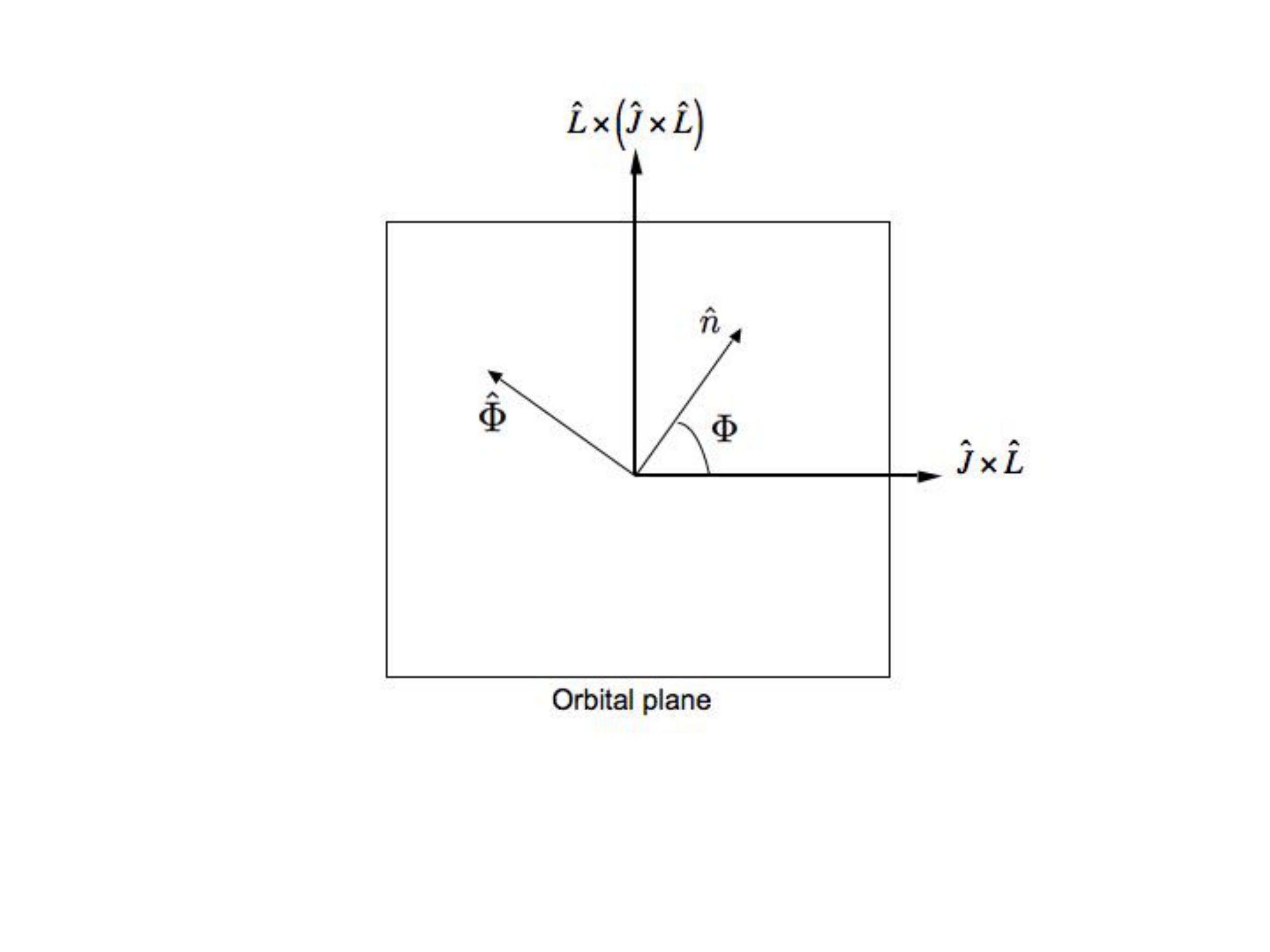}
\hfill
  \caption{The orbital plane coordinates as defined in
    Ref.\ \cite{Levin:2008ci} Fig.\ 1. Left: The orbital plane spanned by
    ${\bf r}\times {\bf p}$ precesses around the total angular
    momentum with frequency $\dot\Psi$. Right: The coordinates as defined
    within the orbital plane.
\label{Fig:OrbitalPlane}
}
\end{figure}

If either of the black holes spin, orbital motion famously no longer
lies in a plane in the absence of special symmetries. The black holes engage in an intricate
three-dimensional motion tangled by precession.  An example of the
three-dimensional range of a
spinning pair is shown in center-of-mass coordinates on the far left of
Fig.\ \ref{Fig:Conservative}. 
We want to highlight the two different precession effects: periastron precession
and precession of the orbital plane. 
The two effects can be cleanly
separated with a natural choice of coordinates \cite{Levin:2008ci,Grossman:2008yk}. 
First, consider precession of the orbital plane.
If the spins of the black hole and the orbital angular
momentum are aligned or anti-aligned, motion will lie in the 
equatorial plane, defined as the plane perpendicular to the total angular
momentum, $\bJ=\bL+\bstot$, where $\bL$ is the orbital
angular momentum and $\bstot$ is the sum of the spins.
However, when the spins are not aligned with the orbital angular
momentum, the spins will precess around the total angular momentum, and
by conservation of the total momentum (i.~e.~$\dot \bJ=0$), the orbital angular momentum
will precess to compensate (i.~e.~$\dot \bL=-\dot \bstot$). Since the 
orbital plane is spanned by $(\br,\bp)$ and is orthogonal to
the orbital angular momentum, ${\bf L}=\br\times \bp$, as shown in
Fig.\ \ref{Fig:OrbitalPlane}, the entire orbital plane
precesses. Plane precession is a purely relativistic reflection of
black hole spins and so isolating the effect draws out distinct
signatures for parameter estimation.

Periastron precession can be isolated, in turn,
by following the motion confined to the orbital plane.
In Ref.\ \cite{Levin:2008ci,Grossman:2008yk}, it was shown that when
there is one effective spin  -- defined by one black
hole spinning or two equal-mass black holes with arbitrary spins -- that periastron and apastron are
constants and the periastron precesses at a fixed rate, as shown in
the middle panel of \ref{Fig:Conservative}. By contrast, these precise
features are obscured in the equatorial plane, as shown by the 
projection in the far right panel of \ref{Fig:Conservative}. So, to
separate periastron precession from the precession of the orbital
plane, we select a coordinate system that cleaves the two effects.

The coordinate system that affects this separation is given by $(r,\Phi,\Psi)$ where $r$ is
the radial coordinate, $\Phi$ is the angle swept out in the orbital
plane, and $\Psi$ is the angle swept out as $\bL$ swings around $\bJ$
\cite{Levin:2008ci,Grossman:2008yk}.
The polar coordinates in the orbital plane that precess through space
as the plane precesses are $(\bn,\bhPhi)$ with
\begin{equation}
\bhPhi=\bhl\times \bn \quad .
\label{Eq:Phihat}
\end{equation}
where $\bn=\br/r$. The entire orbital plane then precesses around the
total momentum in the direction $\bhPsi$ given by
\begin{equation}
\bhPsi=\bhj \times \frac{(\bhj\times \bhl)}{|\bhj\times \bhl|} \quad .
\end{equation}
In the set of coordinates $(r,\Phi,\Psi)$ and their conjugate
momenta $(P_r,P_\Phi,P_\Psi)$, the angular conjugate
momenta are simply $P_\Phi=L$ and $P_\Psi=L_z=\bL\cdot \bhj$. The magnitude of the
orbital angular momentum $L$ is conserved and so, therefore, is $P_\Phi$.

In the restricted case of one effective spin -- again, defined by one black
hole spinning or two equal mass black holes with arbitrary spins -- the component of the
orbital angular momentum along $\bhj$, and therefore $P_\Psi$, is also conserved.  
The Hamiltonian equations of motion then have a
remarkably simple form: 
\begin{align}
\dot r &=\frac{\partial H}{\partial P_r} \ ,  \dot P_r  =
-\frac{\partial H}{\partial r}
\nonumber \\
\dot \Phi &=\frac{\partial H}{\partial P_\Phi} \ ,  \dot P_\Phi  =  0
\nonumber \\
\dot \Psi &=\frac{\partial H}{\partial P_\Psi} \ , \dot P_\Psi  =  0
\label{Eq:eomsimp}
\end{align}
The natural frequencies given by radial oscillations,
$f_r=1/T_r$ where $T_r$ is the radial period,
periastron precession, $f_\Phi=\dot \Phi/2\pi$, and orbital plane precession $f_\Psi=\dot
\Psi/2\pi$, are functions of $r$ only.

The more general equations, when both black holes spin and have unequal masses, and dissipation is included,
are collected in the Appendices. In the coordinate system introduced above,
adding a second spin (for unequal mass) renders $P_\Psi=L_z$ no longer
conserved, although $P_\Phi=L$ continues to be conserved. Consequently, the precessional frequencies will not depend
solely on $r$, but will be 
modulated by angular position. Adding dissipation drains both $P_\Phi$
and $P_\Psi$, although $(r,\Phi,\Psi)$ remains the natural coordinate
system for disentangling the two kinds of precession.

We then convert the radiation-reaction terms derived in the Lagrangian formulation
(\S \ref{sec:lag}) into Hamiltonian variables (\S \ref{sec:ham}). We
are motivated to show our results in the following section in
Hamiltonian variables by the simplicity of Eqs.\ (\ref{Eq:eomsimp})
and the simplicity of the analytic expressions for the frequencies,
as we discuss in the following section (see Eqs.\ (\ref{Eq:noss})).
Of course, practitioners are free to choose either the Lagrangian or
the Hamiltonian formulation,
since both admit equivalent
descriptions, so that an orbital-plane decomposition can be found with
corresponding frequencies. The Hamiltonian formulation  offers a
cleaner framework in which to examine the explicit equations, so we
favor it slightly for the purposes of the paper. \footnote{For example, the
simplicity of 
$\vec L=\vec r\times \vec p$ in the Hamiltonian
formulation renders our geometric interpretation more transparent 
than the lengthy 3PN corrected definition of $\vec L$ in the
Lagrangian variables.} 

\section{Gallery of Inspirals}
\label{sec:gallery}

Using the equations of motion of \S \ref{sec:ham},
we can 
investigate a completely generic orbit. 
The most general scenario
allows for both black holes to spin and for those spins to be
misaligned. 
These three-dimensional orbits should have three natural
frequencies. As described in \S \ref{sec:dynamics}, the natural
frequencies are
the radial oscillations of an
eccentric orbit
\begin{equation}
f_r=\frac{1}{T_r} \quad ,
\label{Eq:fr}
\end{equation}
where $T_r$ is the time between successive periastra,  
the frequency of periastron precession in the orbital plane,
\begin{equation}
f_\Phi=\frac{\dot \Phi}{2\pi} \quad ,
\label{Eq:fPhi}
\end{equation}
and the frequency of plane precession,
\begin{equation}
f_\Psi=\frac{\Dot\Psi}{2\pi} \quad .
\label{Eq:fPsi}
\end{equation}

In the conservative Hamiltonian system, it was found that if spin-spin
contributions were omitted, then \cite{Levin:2008ci}
\begin{align}
2\pi f_\Phi &= \dot \Phi = A\frac{L}{r^2}+\frac{\bs\cdot \bhl}{r^3}-\dot \Psi
(\bhj\cdot\bhl) \nonumber \\
2\pi f_\Psi &=\dot \Psi = \left (\frac{\bhj\times (\bs \times \bhl)}{|\bhj\times
  \bhl| r^3}\right ) \cdot \bhPsi
\label{Eq:noss}
\end{align}
where $A=2\,\partial H/\partial \bp^2$ (Eq.\ (\ref{abcd}),
and $\bs$ is defined in Eq.\ (\ref{Eq:Seff}).
 Although we add spin-spin coupling, we continue to use Eqs.\ (\ref{Eq:noss})
 as adiabatic estimates for the frequencies.

The 
full waveform will be composed of the frequencies (\ref{Eq:fr})-(\ref{Eq:fPsi}).
The waveform is computed to leading quadrupole order, although the orbital variables that go into this expression are
computed in the full 3.5PN system,
\begin{equation}
h_{ij}=\frac{2\mu}{D} 2\left (v_iv_j-\frac{1}{r}\hat r_i\hat r_j  \right )\ \ ,
\end{equation}
with $D$ the
distance to the sources.  This is referred to as the restricted PN waveform approximation.
For simplicity, we place the Earth along $\hat z=\hat J$.
For greater precision, higher-order corrections to the amplitude (including spins) have
been computed \cite{kidder1993,kidder1995}, but for our purposes, the leading-order
amplitude is sufficient. 

\begin{figure}
 \includegraphics[width=0.54\textwidth]{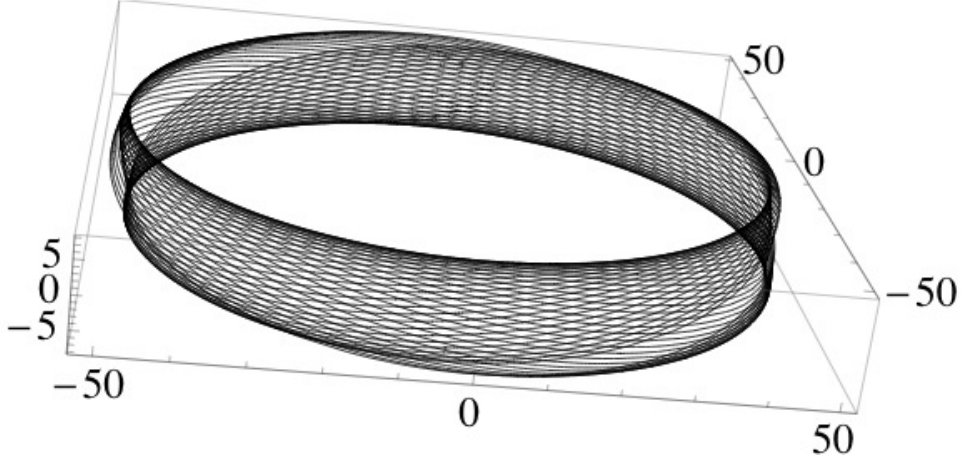}
 \includegraphics[width=0.44\textwidth]{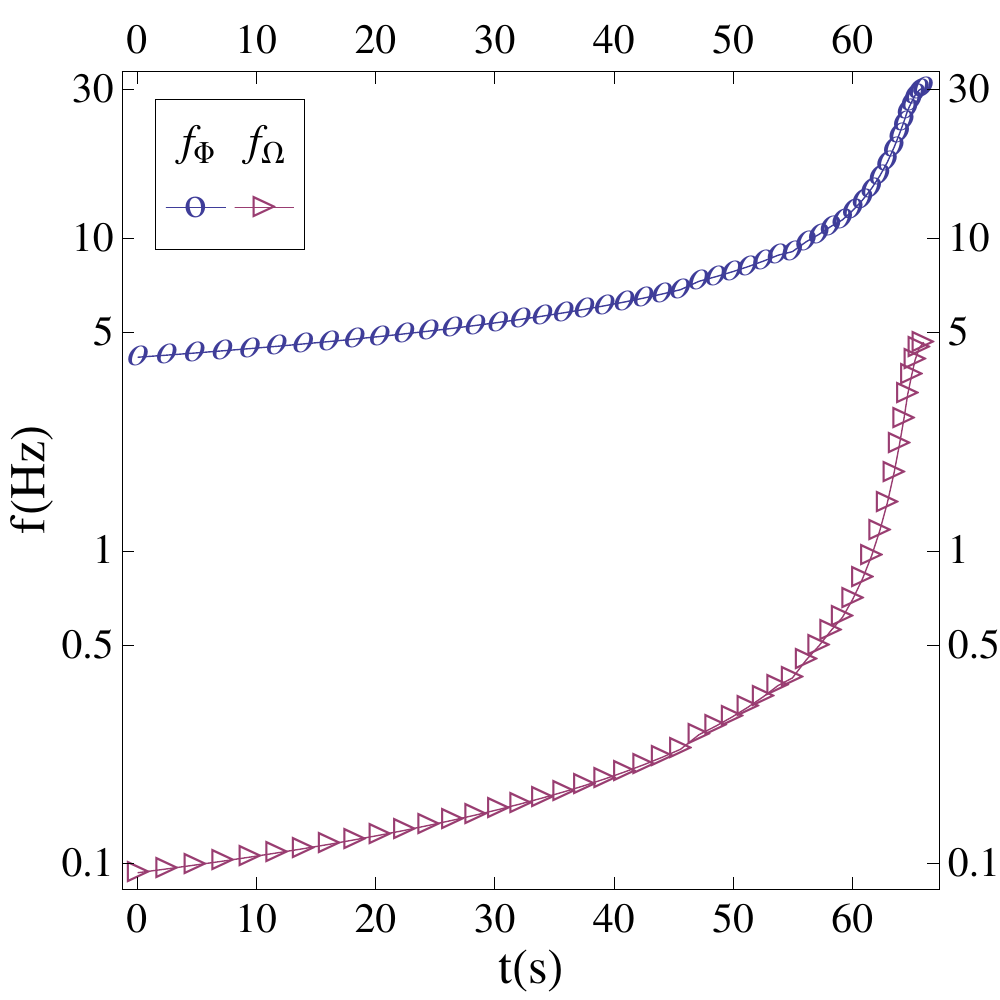}
\hfill
  \caption{The initial conditions are set to
    those of the constant radius orbit at $r_i=50$ in the absence of
    spin-spin coupling for an equal mass binary with 
maximal spins, and spin angles
$\theta_1=\theta_2=45^o$. 
Left: The first 10 sec of the orbit is shown (length
measured in units of $GM/c^2\simeq 26.5\left (M/20M_\odot\right )$km).
Right: The frequency in the
orbital plane as defined in Eq.\ (\ref{Eq:noss}) marked with open
circles,
and the frequency of plane precession, as defined in  Eq.\ (\ref{Eq:noss}),
marked with crosses. Both are shown versus time over the
entire inspiral.
We plot frequencies
in units of $\left (20M_\odot/M\right )$Hz and time in
units of $(M/20M_\odot)$sec. 
\label{Fig:QuasiSphere1}
}
\end{figure}

\subsection{Quasi-spherical}
\label{sec:qs}

We now consider 
quasi-spherical, low eccentricity orbits
(see also  \cite{Buonanno:2005xu}). Low eccentricity orbits look
quasi-circular in the orbital plane. As the orbital plane
precesses, these orbits
lie on the surface of a sphere whose radius decreases monotonically as
gravitational radiation is emitted.
We work in the Hamiltonian formulation, since the Hamiltonian better
lends itself to finding constant radius orbits and facilitates the
geometric breakdown in terms of the orbital plane. Of course, this
procedure can readily be repeated in the Lagrangian formulation.
We find the constant radius orbit according to the
prescription in \cite{Levin:2008ci} (see also
\cite{Buonanno:2005xu}). We sketch the argument here. The
Hamiltonian of Eqs.\ (\ref{Eq:HPN}) and (\ref{Hterms}) does not admit
a simple
effective potential formulation since it is a complicated function of
$\bp^2$. Nonetheless, we can still use the Hamiltonian as an effective
potential {\it at the turning points} if we exclude spin-spin coupling: 
\begin{equation}
V_{\rm eff}=H({P_r=0}) \quad ,
\label{Eq:Veff}
\end{equation} 
With this condition, we can find the $L$ of an orbit at a given
constant radius.
Fixing the initial conditions this way and including
spin-spin couplings, we can estimate the deviation away from quasi-sphericity.
We find that the oscillations in
the radius due to spin-spin coupling are negligible (less than a few
percent) and that the radius quickly begins a monotonic decrease as
gravitational waves are radiated. Therefore, the spin-spin effect
seems to be
too small to induce measurable eccentricity during the inspiral if no eccentricity is present initially.

As an example, consider an equal mass pair 
with maximal spins
 $a_1=a_2=1$, and spin angles initially misaligned so that
$\theta_1\equiv \arccos({\bf\hat L}\cdot{\bf\hat S_1})=45^o$ and
$\theta_2\equiv \arccos({\bf\hat L}\cdot{\bf\hat S_2})=45^o$.  This is  shown in 
Fig.\ \ref{Fig:QuasiSphere1}, where the first ten seconds of the total orbit is shown. 
The full orbit takes $\sim 60$ seconds to merge for
$M=20M_\odot$.
We take the initial condition to be
$r_i=50$ and set $L$ initially at the quasi-spherical value 
determined by Eq.\ (\ref{Eq:Veff}).
The
natural frequencies $f_\Phi$ and $f_\Psi$ are shown in the right panel. Since the
orbit is quasi-spherical, the frequency of
radial oscillations $f_r$ is not informative. Periastron precession dominates over plane precession,
$f_\Phi > f_\Psi$, which is expected since plane precession is a higher-order effect.

Using physical units,
with the initial separation of $r_i\sim 50$, the 
gravitational radiation has an initial frequency 
\begin{equation}
2\times f_\Phi \sim 9.5{\rm Hz}\left (\frac{20M_\odot}{M}\right )
\end{equation}
which is already nearing the Advanced LIGO/VIRGO band for two $10 M_\odot$ black
holes, and is sweeping through the bandwidth as the orbital separation decreases.
The much slower
frequency of precession, $f_\Psi$, does not have a significant effect on the
waveform for the quasi-spherical case.

\begin{figure}
\includegraphics[width=0.49\textwidth]{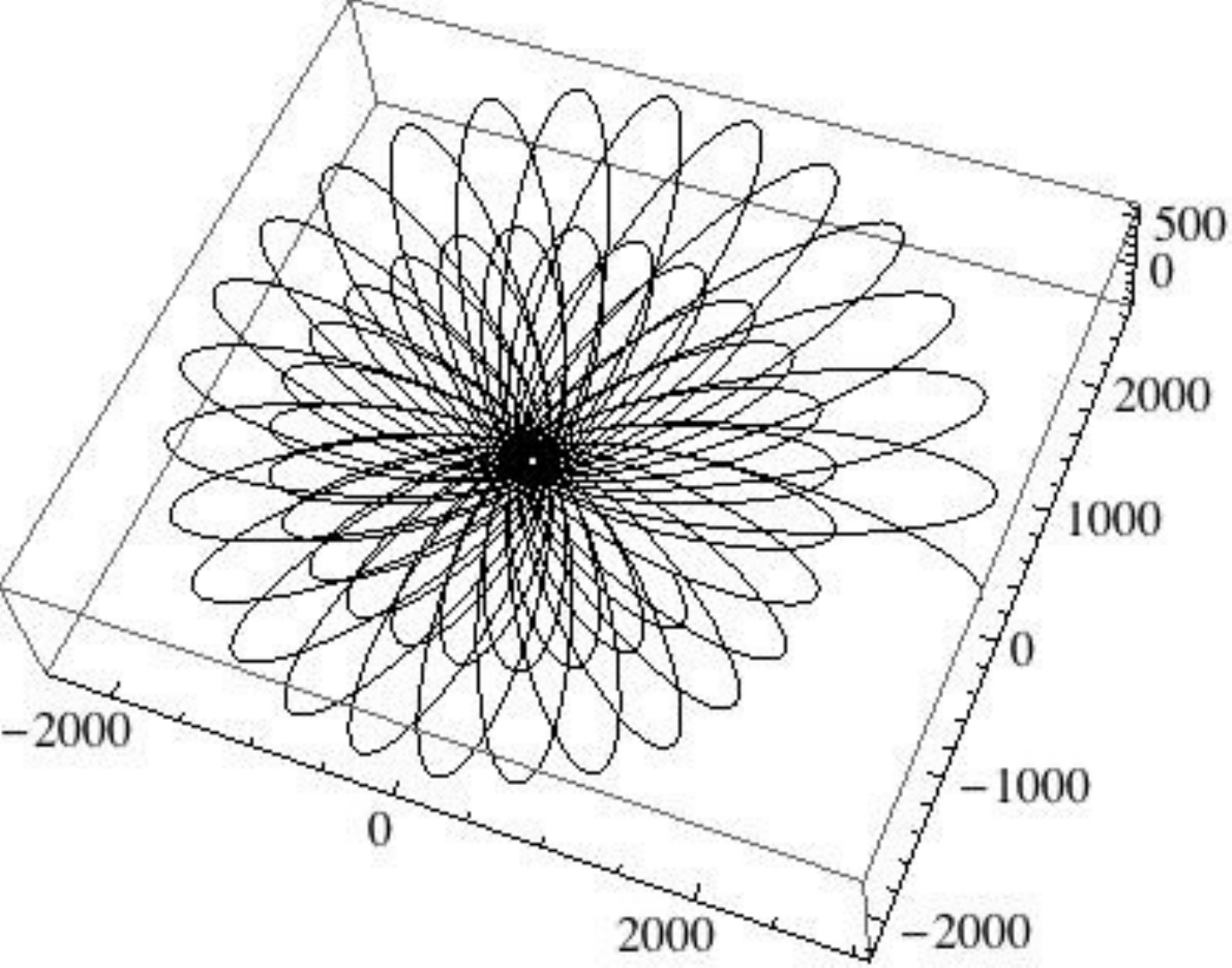}
\centering
\includegraphics[width=0.49\textwidth]{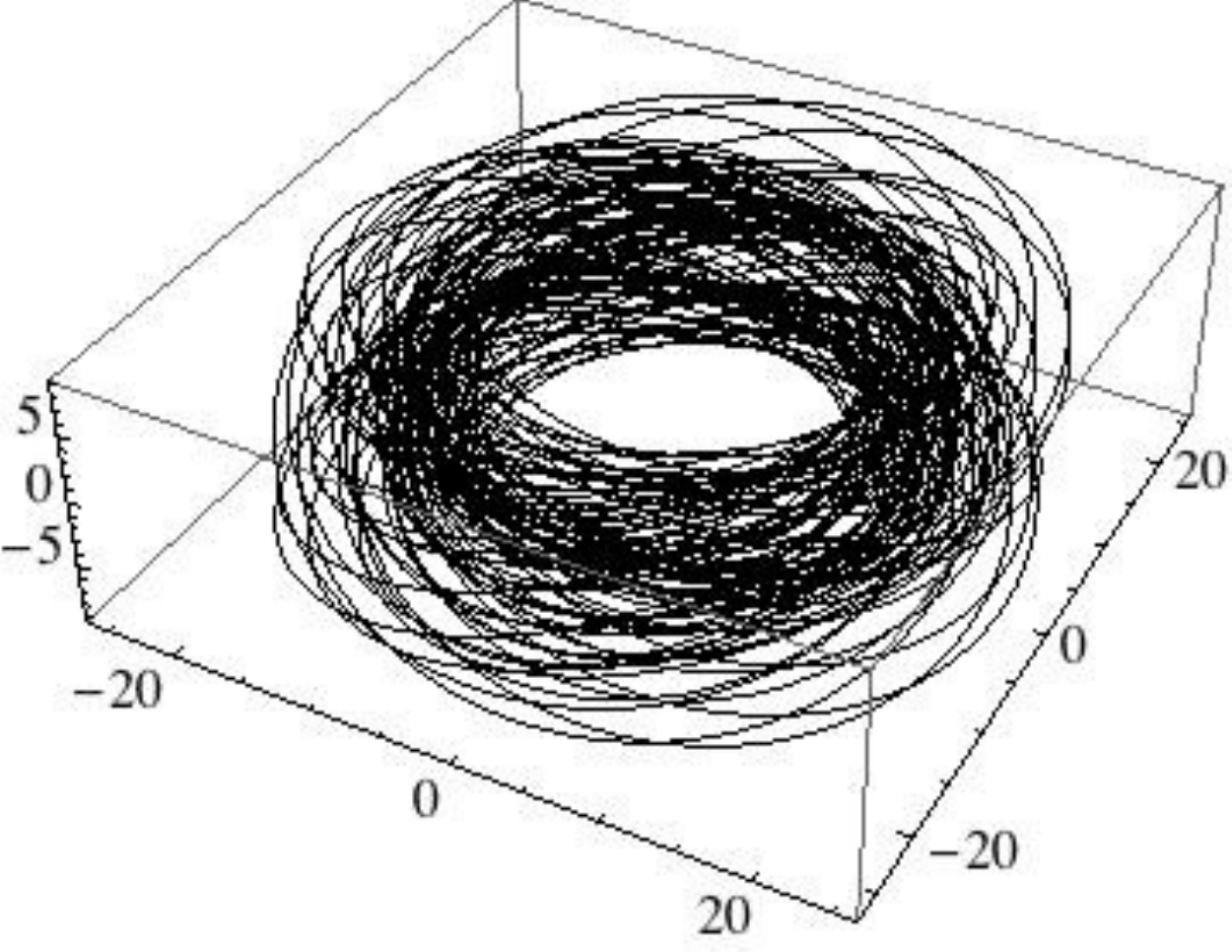}
\hfill
  \caption{The pair has $m_2/m_1=1/4$, maximal spins,
$a_1=a_2=1$, and initial spin angles
$\theta_1=45^o,\theta_2=45^o$ and initial values $r_i=3000$ and $L_i=0.003
    r_i$. Left: The first 1150s of the inspiral.
   Right: The final 5s before cutoff.
  \label{Fig:uneqburstyorb}
}
\end{figure}

\begin{figure}
\includegraphics[width=0.32\textwidth]{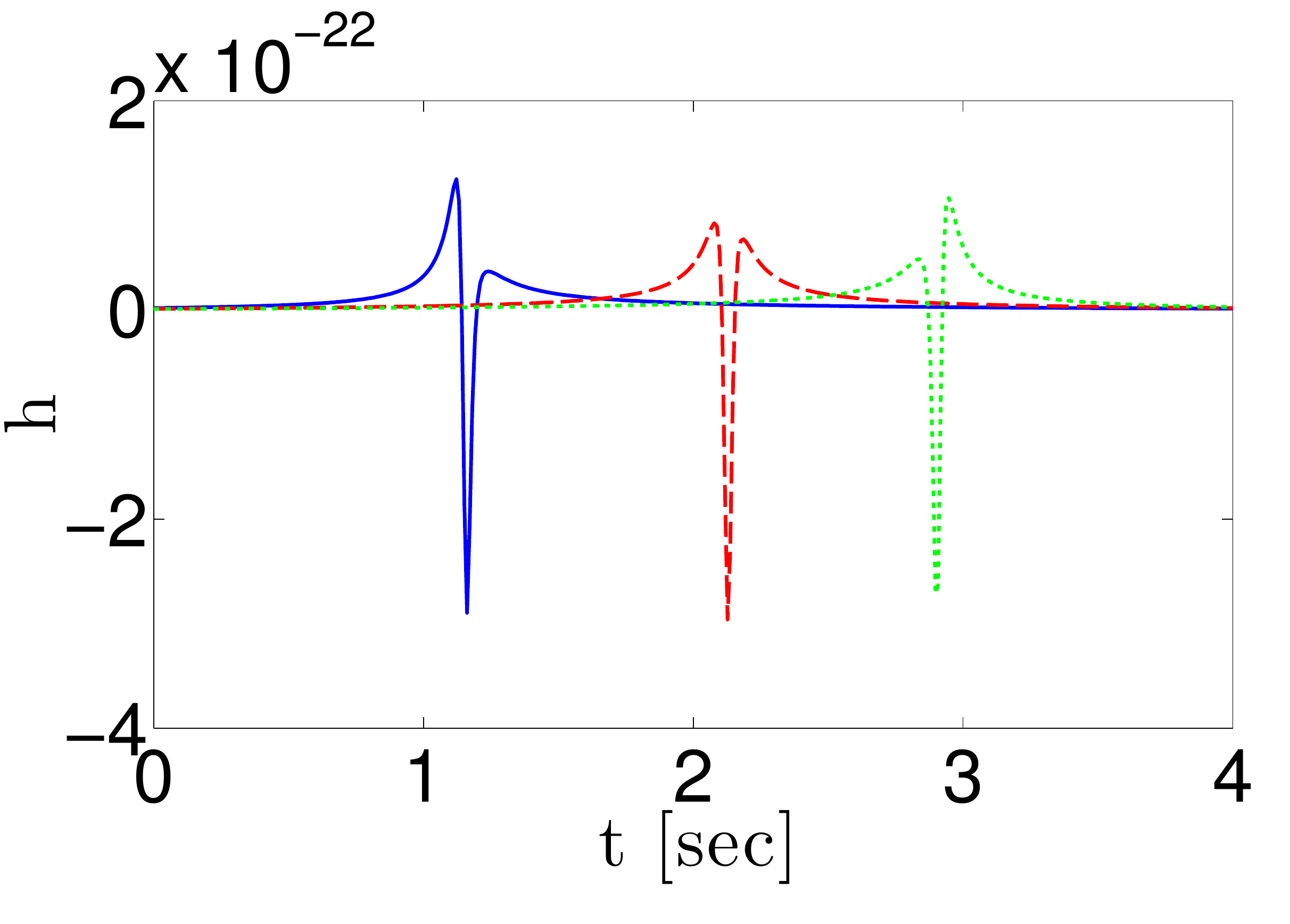}
\includegraphics[width=0.32\textwidth]{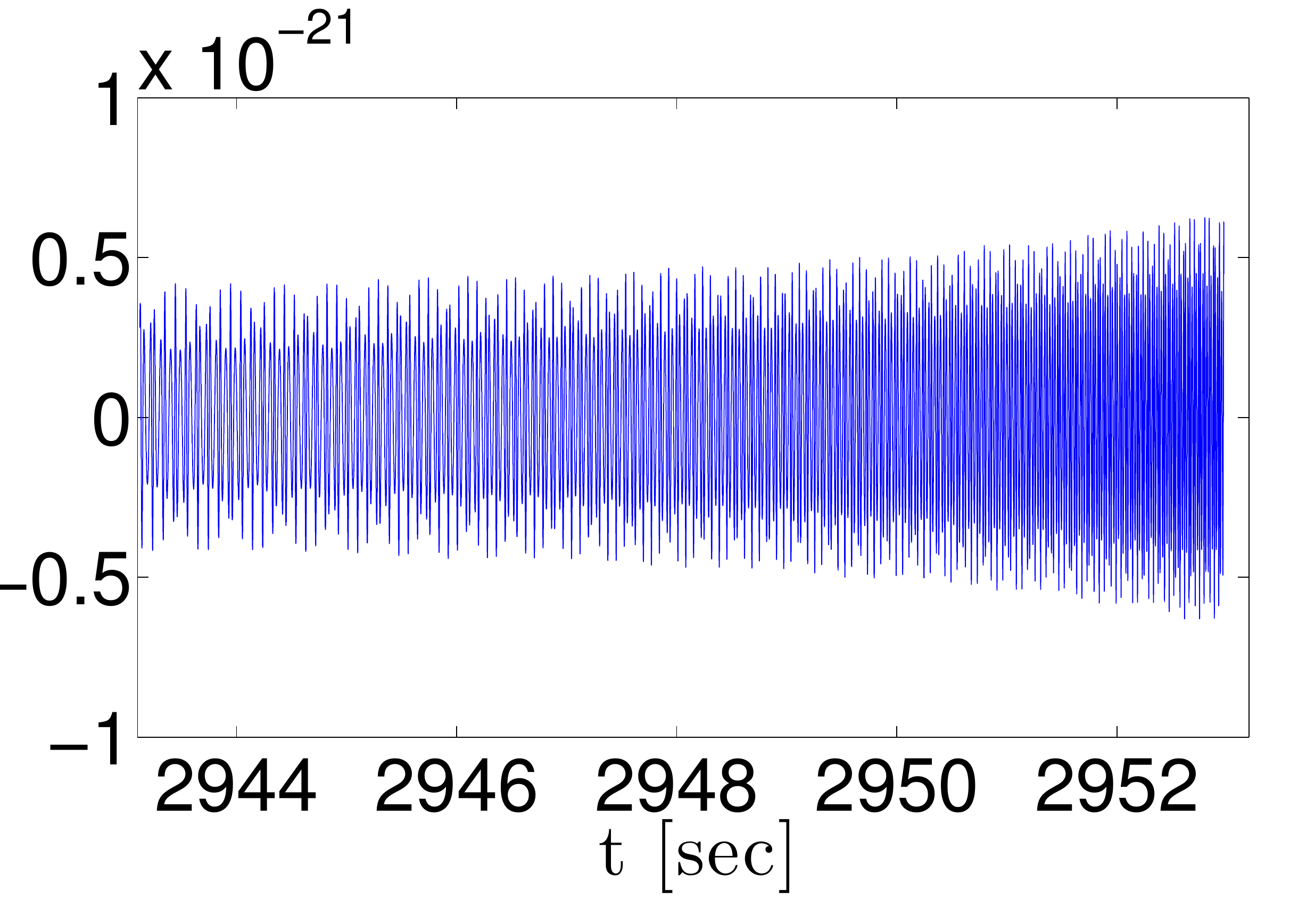}
\includegraphics[width=0.32\textwidth]{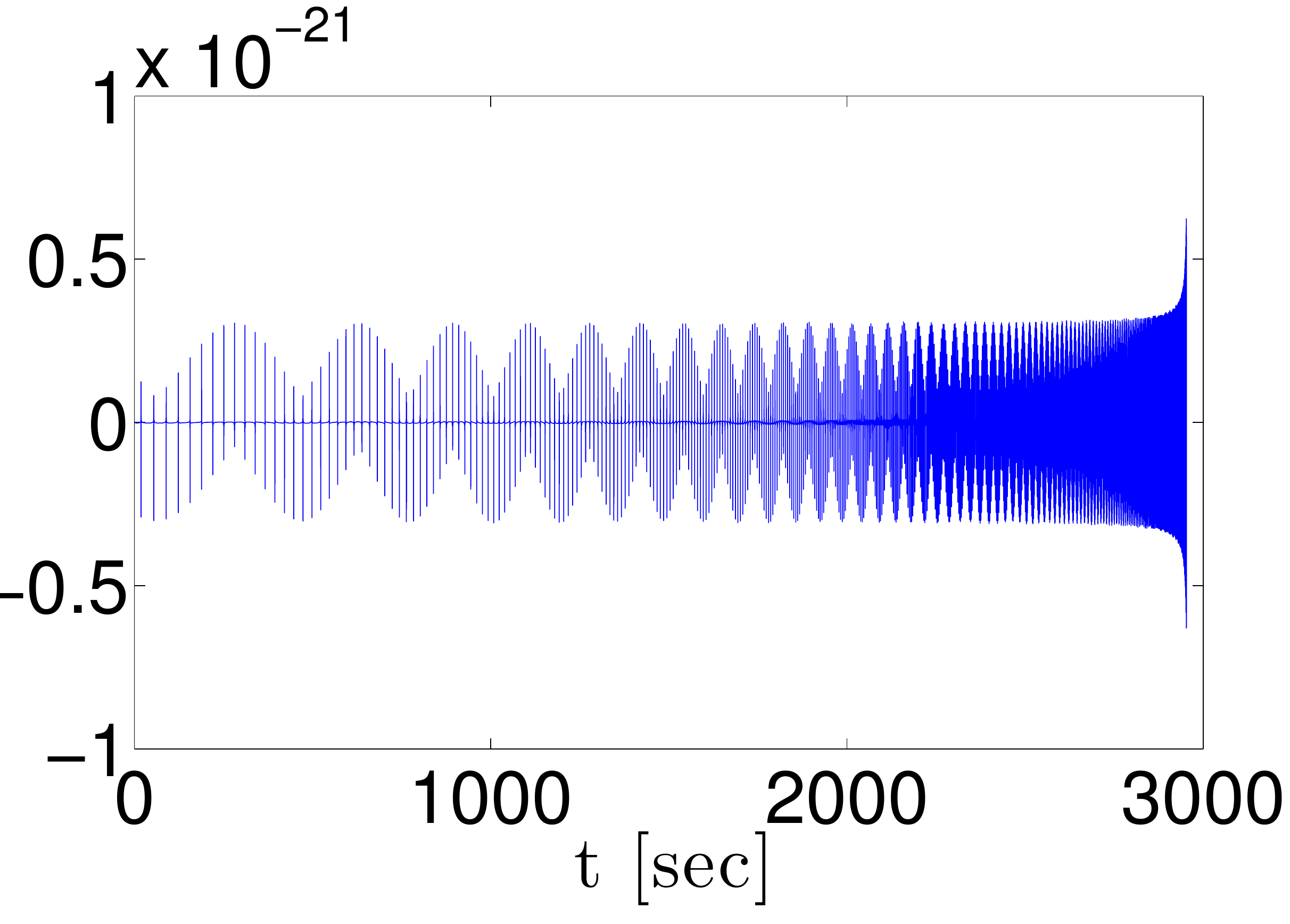}
\hfill
\caption{Strain for an optimally-oriented binary with $M=20 M_\odot$ at $D=100$ Mpc.
From left to right, we show three periastron passes occurring $\sim3000$ sec prior to waveform truncation
(shifted in time by $\sim34$ sec each to increase their overlap), 
the final 10 seconds, and the final $\sim3000$ seconds of the waveform.
\label{Fig:uneqburstywave}
}
\end{figure}

\begin{figure}
\includegraphics[width=0.4\textwidth]{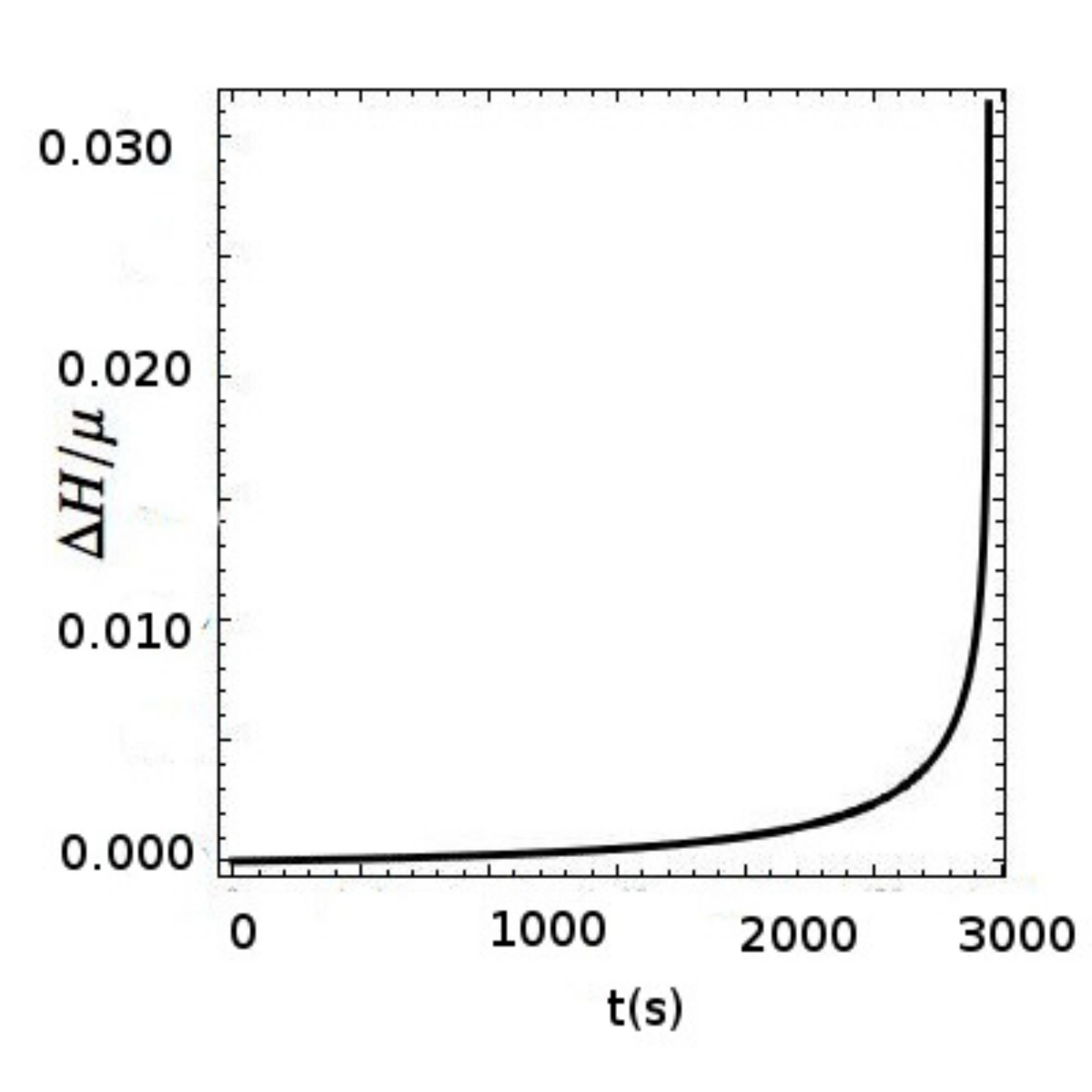}
\includegraphics[width=0.4\textwidth]{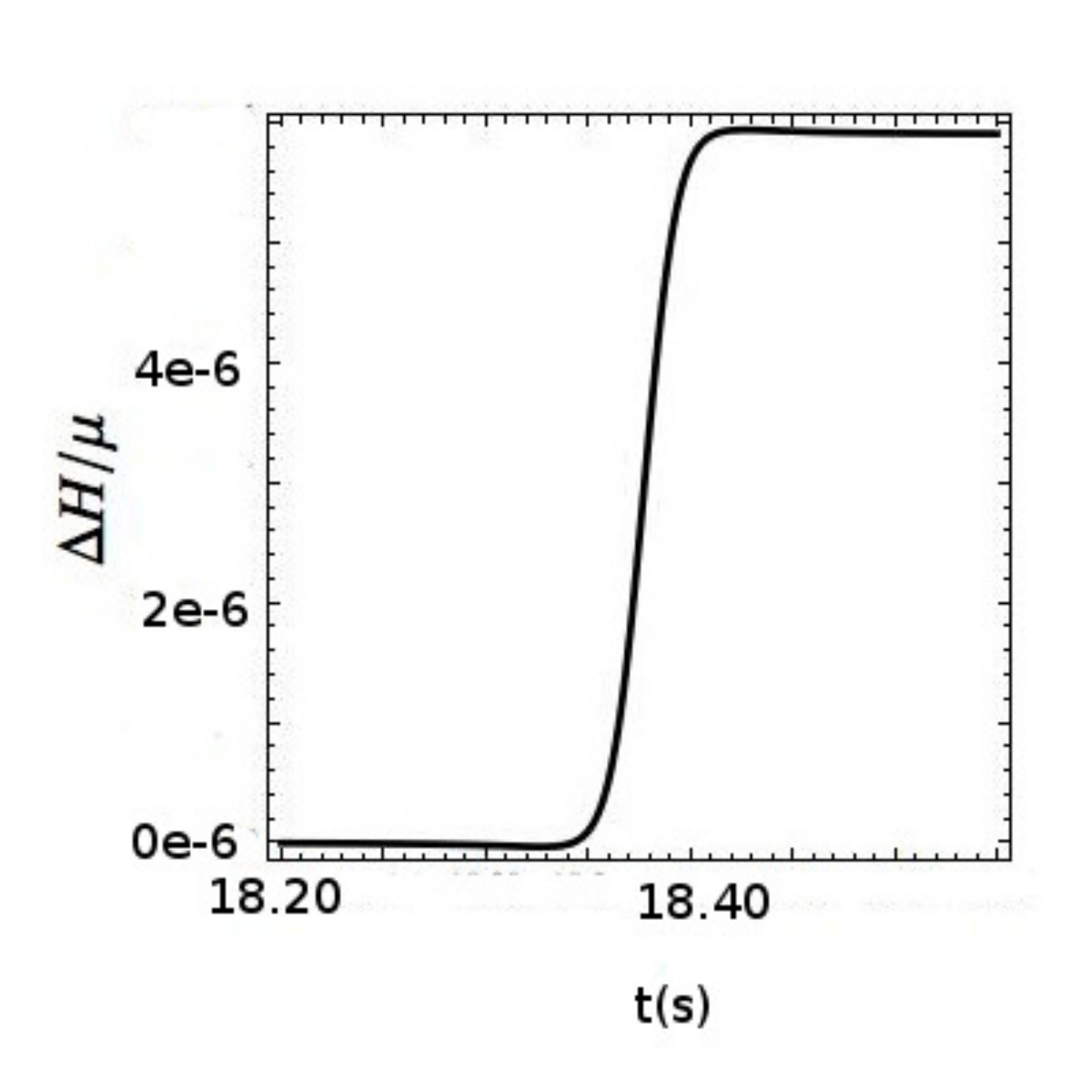}
\hfill
\caption{$\Delta H$ is the absolute value of the
difference between the energy as a function of time and the initial energy.
Left: The burst of energy emitted during one periastron passage
($r_p\sim 38$). As the pair separate to apastron $r_a\sim 3000$, negligible energy is
lost as confirmed by the flatness of $\Delta H$.
Each burst is small although a few percent of the total
energy in $\mu$ is lost cumulatively prior to cutoff as shown on the
right.
\label{Fig:uneqbursty}}
\end{figure}

\begin{figure}
\includegraphics[trim = 0mm 0mm 0mm 20mm, width=0.44\textwidth]{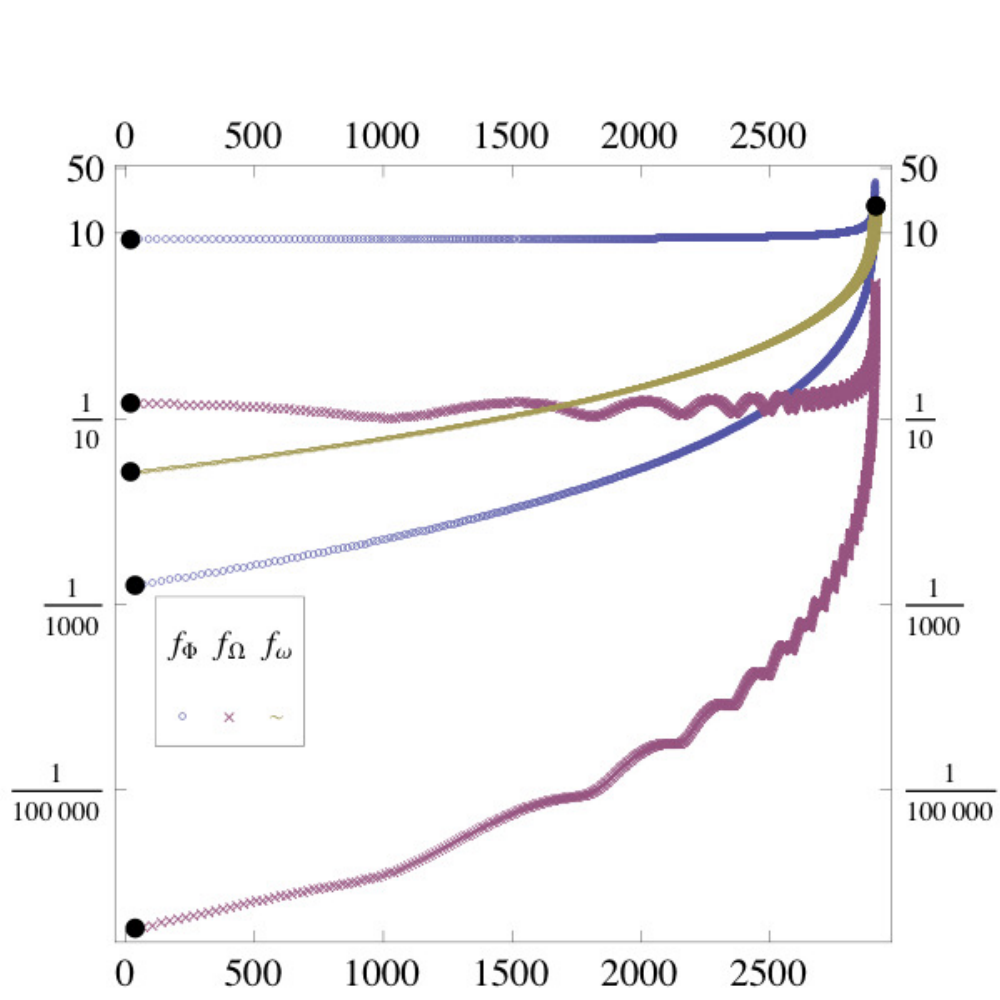}
\includegraphics[width=0.54\textwidth]{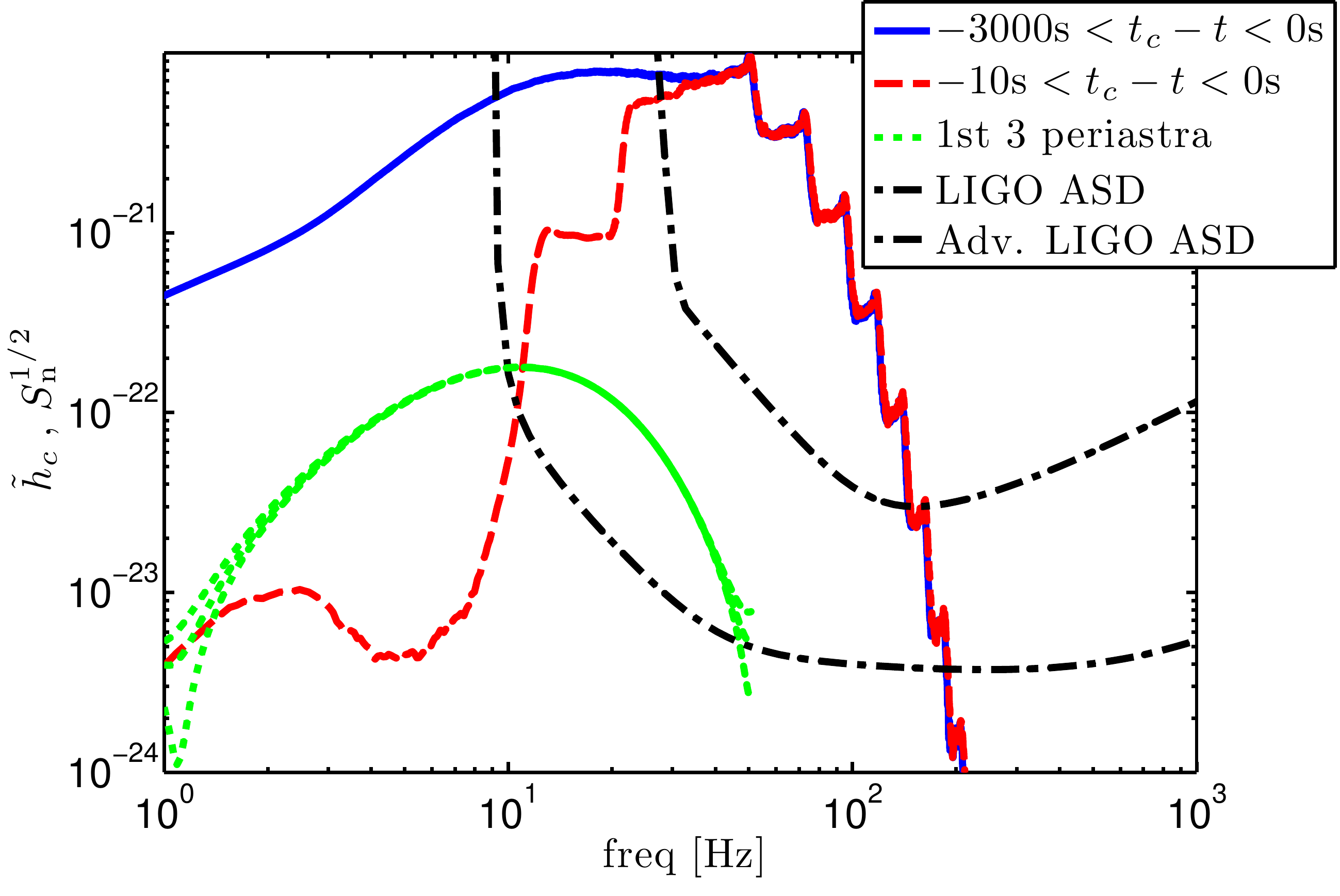}
\centering
\hfill
\caption{
Left: The three natural frequencies versus time for the eccentric configuration. There are two lines for $f_\Phi$, one
marking the value at periastron (top line) and the other marking the value at
apastron (bottom line). Same for
$f_\Psi$.
All frequencies are plotted
in units of $\left (20M_\odot/M\right )$Hz and all times
in units of $(M/20M_\odot)$sec. 
Right: The characteristic strain, $h_c$, of an optimally-oriented realization of the eccentric configuration, at a distance of
100 Mpc, again scaled for $M=20M_\odot$.  We show $h_c$ for each of the first three periastron passages, the final 100 sec, and
the entire 3000 sec intervals shown in Fig.~\ref{Fig:uneqburstywave}.  We show the LIGO and Advanced LIGO sensitivity curves \cite{OShaughnessy}
for comparison.
\label{Fig:uneqburstyFT}
}
\end{figure}

\subsection{Eccentric, unequal masses}
\label{sec:ecc}

We next consider an eccentric pair
with
mass ratio $m_1/m_2=1/4$.
Each black hole is spinning maximally. Both spins
are misaligned with the orbital plane with 
$\theta_1\equiv \arccos({\bf\hat L}\cdot{\bf\hat S_1})=45^o$ and
$\theta_2\equiv \arccos({\bf\hat L}\cdot{\bf\hat S_2})=45^o$.
The pair
begins at large separation, $r_i=3000$, and with angular momentum
$L=0.003 r_i$. The first periastron passage is $r_p\sim 38$, so the
pair is set on a highly eccentric orbit.
The eccentricity can be defined loosely as
\begin{equation}
e=\frac{r_a-r_p}{r_a+r_p}
\end{equation}
where $r_a$ is the apastron and $r_p$ is the periastron. The eccentricity
 begins large, $e=0.97$, and
drifts down to $e=0.17$ by the time the simulation is ended. Cutoff is set when the 3.5PN terms become larger
than the 2.5PN terms, or, equivalently,
when $\dot E>0$, as we discuss in the next section. At cutoff, the
periastron is $r_p\sim 12$ and the apastron is
$r_a\sim 17$.
Our choice of the initial pericenter distance and eccentricity
is consistent with the capture formation scenario of stellar mass black hole binaries in globular clusters \cite{O'Leary:2008xt}.
Therefore, an initially highly eccentric orbit could retain significant eccentricity by the time it evolves into the bandwidth of
a terrestrial network of gravitational-wave interferometers.

The pair
experiences both periastron precession and orbital-plane precession
during a total of 1384 orbits, some of which are shown in Fig.\ \ref{Fig:uneqburstyorb}.
On the left of Fig.\
\ref{Fig:uneqburstyorb} 
the first 1150s of the orbit are shown and on the right the final 5s
of the orbit is shown before the simulation is cutoff.
Both the precession of the periastron and the precession of the
orbital plane are apparent in
    Fig.\ \ref{Fig:uneqburstyorb}.
The precession of the orbital plane is more prominent the more
    disparate the masses of the black holes.
While spin precession remains substantial to the end, periastron precession
becomes less significant as eccentricity is shed.

Since the orbit is highly eccentric, a burst of
gravitational radiation is emitted near periastron passage as shown in
Fig.\ \ref{Fig:uneqburstywave}.
During the most eccentric cycles, the bursts are what one might expect
from a parabolic encounter, with 
energy and angular
momentum losses occurring primarily near periastron.
A tiny amount of energy is emitted per periastron passage for the
$M=20M_\odot$ pair shown in Fig.\ \ref{Fig:uneqbursty}, although a few percent of the total $\mu$ is
shed prior to plunge. 
The bursty waveform 
is highly distinct from the extremely
regular quasi-circular
waveforms, clearly showing effects of the precession.

The natural frequencies $f_\Phi>f_r>f_\Psi$ are shown on the left of 
Fig.\ \ref{Fig:uneqburstyFT}. Both $f_\Phi$ and $f_\Psi$ are shown
only at periastron (top line) and
apastron (bottom). Therefore, there are two lines each for $f_\Phi$ and $f_\Psi$. The
value of the frequencies at apastron is much smaller than at periastron
if the eccentricity is large. These discrete values approach each other
as the eccentricity decreases. Since $T_r$ is defined as the time
between successive perihelia,
there is only one line for
$f_r=1/T_r$.
The wobbling in $f_\Psi$ is a result of spin-spin coupling.
The time between bursts is given by the radial
period, $T_r=1/f_r$, and the frequency of the burst is given
by $f_\Phi$ at periastron passage during the highly-eccentric, nearly parabolic
encounter, and by 
$2\times f_\Phi$ when the orbit is less eccentric.

The right panel of 
Fig.\ \ref{Fig:uneqburstyFT} shows the characteristic strain,
\begin{equation}
h_c \equiv 2 f \sqrt{\left|\tilde{h}_+\right|^2 + \left|\tilde{h}_x\right|^2}\,,
\end{equation}
for the three time intervals shown in Fig.\ \ref{Fig:uneqburstywave}, as well
as the noise amplitude spectral densities for the initial and Advanced LIGO detectors.
The self-similarity of the periastron passages is evident in Fig.\ \ref{Fig:uneqburstyFT}.
We suggest the possibility that this self-similarity could be employed to coherently combine the signals from all the periastron
passages of a highly eccentric system, while ignoring the data in between where the signal power is negligible.

The characteristic strain of the full signal has a flat plateau that spans the frequency
range between the spin precession frequency and twice the orbital frequency, both evaluated at periastron,
because the power emitted during each orbit is dominated by the emission at periastron.  The characteristic
strain of the final 10 sec shows a very clear step structure due to the large set of harmonics excited by the eccentric motion.
We emphasize that these harmonics are not the instantaneous harmonics from PN corrections, as we have not included higher-harmonic PN corrections. 
Rather, the harmonics apparent in Fig.\ \ref{Fig:uneqburstyFT} are the result of the large change in instantaneous frequency
over the course of each orbit, which is represented in Fourier space as higher (and lower) harmonics of twice the mean orbital frequency.
Again, the highest plateau spans between the spin precession frequency and twice the orbital frequency, with the step at lower
frequency occurring at the orbital frequency, and the many steps at higher frequencies occurring at larger integer multiples
of the orbital frequency.

\section{Domain of validity for the PN approximation}
\label{sec:valid}

The PN equations of motion are a series expansion of corrections to the acceleration of the binary components,
with whole order terms representing
conservative, relativistic corrections, and half order terms representing
dissipative corrections.  As such, the expansion should not be trusted when higher-order terms become comparable in magnitude
to lower-order terms.
In this regard, it is noteworthy that the sign of the acceleration term along the radial
direction is negative at 2.5PN order,
as expected, and the black holes are drawn together. However, the sign of the
acceleration term along the radial direction is positive at 3.5PN
order. In other words, the 2.5PN terms over estimate the effects of
dissipation, and the 3.5PN terms temper this over estimate. A
consequence of the breakdown of the PN expansion in the strong-field regime is that
the 3.5PN term comes to dominate, and spuriously drives the system to a larger separation, with
the 3.5PN term acting
as a source, rather than a sink, of energy and angular momentum.
Implementation of the equations of motion must be
cut off before this happens, as the PN approximation has already broken down at this point.
We find that the break down tends to happen at radial separations $r \sim 10$ in units of the
total mass, which is well outside the Schwarzschild ISCO at $6M$ that is frequently used to represent the
boundary for a valid application of the PN expansion.

\begin{figure}
\centering
\includegraphics[width=0.85\textwidth]{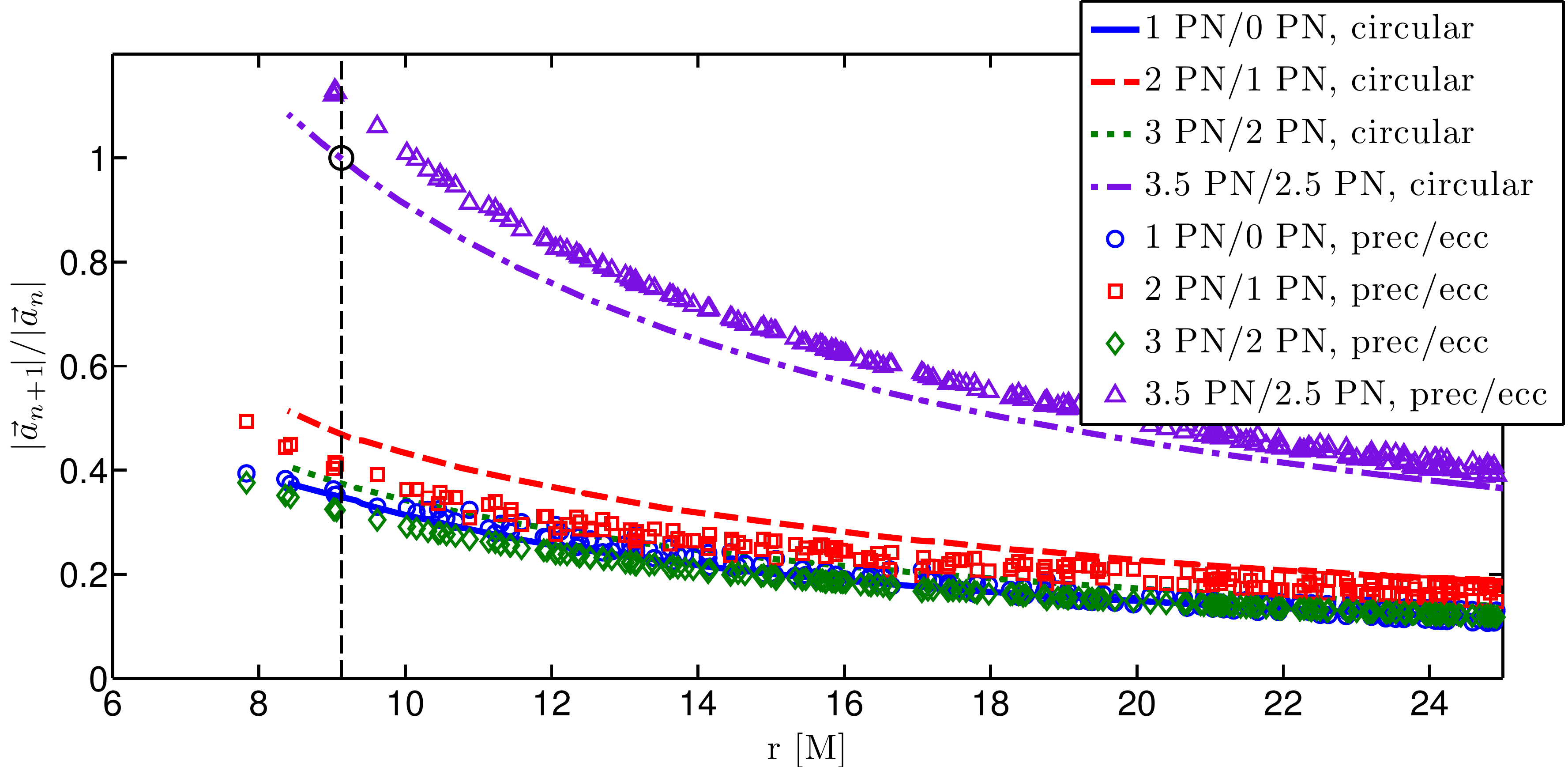}
\hfill
\caption{
The ratio of different PN-order contributions to the Lagrangian equations of motion for a circular, equal mass, nonspinning inspiral, and
separately for our eccentric configuration.  Even for the circular case, the 3.5 PN dissipation term grows larger than the 2.5 PN
dissipation term for radial separations less than $\sim9M$.  Less ideal cases, not surprisingly, appear to make the PN sequence
diverge at larger radii.  
\label{Fig:pncomp}
}
\end{figure}

All of the simulations in the preceding section are therefore cut off when the PN approximation begins to
break down.  Specifically, we use the criterion that the simulation ends
when the rate of energy loss drops below some threshold, indicating that the 3.5PN correction is approaching the same magnitude as
the 2.5PN term.  In Fig.~\ref{Fig:pncomp}, we show the relative
magnitude of different PN-order contributions to the (Lagrangian) equations of motion.
We focus on two cases: an equal mass, nonspinning binary undergoing quasi-circular inspiral, and our eccentric configuration
from the previous section.  The first case is likely to be a best-case scenario for the convergence of the PN expansion,
while the latter is likely to over extend the PN expansion at larger orbital separations. 
We find that, for the conservative terms, the ratio of adjacent whole-order terms in the PN sequence remains less than
unity over the full domain, and indeed would remain so down to the aforementioned ISCO radius.  However,
the ratio of the two half order, dissipative terms exceeds unity at $r=9.125M$ in the circular case, and at $r=10M$ in the eccentric spinning case.
Since the approximation is certainly unreliable prior to the point where this ratio equals unity, we suggest that generally
the PN expansion cannot be reliably applied for separations $r\lsim 10M$ for any configuration when dissipation is included.

Another possibility is that the optimal asymptotic expansion for the PN sequence should be truncated at 3 PN order.  This is an interesting
possibility, and would in fact be consistent with findings in the Schwarzschild test-mass limit, where much higher PN terms are available
to demonstrate this behavior \cite{YunesBerti}.  Since for comparable masses we lack higher PN-order terms, it is not yet possible to distinguish 
between these two possibilities.

\section{Summary}
\label{sec:summary}

We have compiled the equations of motion of
Refs.\ \cite{Mora:2003wt,Pati:2002ux,Will:2005sn,Wang:2007ntb}, and have
transformed them into Hamiltonian coordinates for ready-to-use equations
governing general black hole binaries.
With these in hand, we can
study any black hole pair, including the effects of spin, eccentricity, and their
accompanying precessions. Given this flexible system, we have answered four
basic questions:

\begin{itemize}

\item `Do spinning pairs tend to quasi-spherical orbits?'

We find that the deviation for a constant radius orbit due to
spin-spin coupling is insignificantly small, so quasi-spherical
orbits are accessible. Consequently spinning pairs do tend to
quasi-spherical orbits.

\item `What features are generically introduced into waveforms through periastron and spin precession?'

As was shown in
\cite{Levin:2008ci}, 
an orbit can be projected onto an orbital plane. The characteristic
frequency within the plane supplants the usual coordinate frequency.
The frequency of plane precession and the frequency of radial
oscillations provide the other crucial
frequencies for analysis.  We have shown that these three frequencies leave a characteristic
imprint on the waveforms in Fourier space.
Furthermore, the frequency of bursts for highly eccentric orbits and the quiescent time
between bursts directly reflect these natural harmonics and
could encourage novel data analysis techniques.

\item `How much
energy is lost during each burst near periastron passage?'

Given the caveat that most of the energy will be lost during the merger, which is beyond the reach of the PN approximation, 
for highly eccentric binaries we show that less than $10^{-3}$ of
a percent of the reduced mass $\mu$ is lost to gravitational radiation
per periastron passage. Some pairs may execute thousands of orbits
before plunge so that a few percent of $\mu$ is lost during inspiral up
to $r\sim 10M$.

\item `How much of the orbit and waveform for eccentric, precessing orbits
is well described by the PN expansion?'

The PN expansion 
has broken down whenever higher-order corrections become larger than lower-order corrections.
We find this tends to happen
at $r \sim 10M$, which is well outside of the Schwarzschild
ISCO that is often used to demarcate the breaking point of the PN approximation.
Possibly the PN sequence diverges inside $\sim 10M$,
or it may be that the optimal PN expansion inside this radius should be truncated at no higher than 3 PN order.

\end{itemize}

The orbits shown give a sample of the complete range that can be
probed in detail for generic, spinning, precessing, eccentric black
hole pairs. We intend to build a data set of black hole pairs to be
made available as a testing ground for developing data analysis
techniques. 

\bigskip
\bigskip
\bigskip

*Acknowledgements*

We are grateful to Clifford Will for his invaluable insights and to
Jameson Rollins for important conversations.
This work was supported by an NSF grant AST-0908365. 
JL gratefully acknowledges
support of a KITP Scholarship, 
under Grant no. NSF PHY05-51164.

\appendix
\section{Lagrangian Equations of Motion}
\label{sec:lag}

In this section we compile the equations of motion including spin
couplings and dissipation as computed by Will
and collaborators over the series of papers \cite{Mora:2003wt,Pati:2002ux,Will:2005sn,Wang:2007ntb}.
We measure length in units of total mass $M=m_1+m_2$ and write the equations in
terms of the dimensionless center-of-mass coordinate $\bx$ and
center-of-mass velocity $\bv$. The reduced mass is defined
as $\eta=\mu/M = m_1m_2/M^2$.
The Lagrangian equations of motion in $\bx,\bv$ with spins added are 
(with $x=\sqrt{\bx\cdot\bx}$ the harmonic radial coordinate
and $\dot x=\bn\cdot \bv$) as compiled from Refs.\
\cite{Mora:2003wt,Pati:2002ux,{kidder1995},{kidder1993}} 
\begin{eqnarray}
\label{keom1}
\dot \bx &=& \bv \\
\ddot \bx &=& {\bf a}_N+{\bf a}_{1PN}+{\bf a}_{2PN}+{\bf
  a}_{2.5PN}+{\bf a}_{3PN}+{\bf a}_{3.5PN}+{\bf a}_{PN-SO} +{\bf
  a}_{3.5PN-SO}
+{\bf a}_{PN-SS} +{\bf a}_{3.5PN-SS}\nonumber 
\end{eqnarray}
and
\begin{align}
\label{keom2}
{\bf a}_N &=-\frac{\bn}{x^2}\nonumber \\
{\bf a}_{1PN} &=-\frac{\bn}{x^2}\left
\{(1+3\eta)\bv^2-2(2+\eta)\frac{1}{x}-\frac{3}{2}\eta\dot x^2\right
\}+\frac{\bv}{x^2}2(2-\eta)\dot x\nonumber \\
{\bf a}_{2PN}&=-\frac{\bn}{x^2}\left\{\frac{3}{4}(12+29\eta)\frac{1}{x^2}+\eta(3-4\eta)(\bv^2)^2+\frac{15}{8}\eta(1-3\eta)\dot
x^4\nonumber\right. \\
 & {}-\left.\frac{3}{2}\eta(3-4\eta)\bv^2\dot
x^2-\frac{1}{2}\eta(13-4\eta)\frac{\bv^2}{x}-(2+25\eta+2\eta^2)\frac{\dot
  x^2}{x}\right \}\nonumber \\
&+\frac{\bv}{x^2}\left \{\frac{\dot x}{2}\left
[\eta(15+4\eta)\bv^2-(4+41\eta+8\eta^2)\frac{1}{x}-3\eta(3+2\eta)\dot
  x^2\right ]\right \} \quad .\nonumber\\
{\bf a}_{2.5PN}&=\frac{\bn}{x^2}\left\{
\frac{8}{5}\eta\left (\frac{1}{x}\right )\dot x\left
(\frac{17}{3}\frac{1}{x}+3v^2\right )
\right \}\nonumber \\
&-\frac{\bv}{x^2}\left \{
\frac{8}{5}\eta\left (\frac{1}{x}\right )\left (3\frac{1}{x}+v^2\right )
\right \} \quad .\nonumber
\end{align}
\begin{align}
{\bf a}_{3PN}&=-\frac{\bn}{x^2}\left\{
-\left [16+\left (\frac{1399}{12}-\frac{41}{16}\pi^2\right )\eta +\frac{71}{2}\eta^2
\right ]\left (\frac{1}{x}\right )^3
-\eta\left [\frac{20827}{840}+\frac{123}{64}\pi^2-\eta^2\right ]
\left(\frac{1}{x}\right )^2v^2
\right. \nonumber \\
&\left.
+\left [1+\left(\frac{22717}{168}+\frac{615}{64}\pi^2\right
  )\eta+\frac{11}{8}\eta^2-7\eta^3\right ]
\left (\frac{1}{x}\right )^2\dot x^2
\right. \nonumber \\
&\left.
+\frac{\eta}{4}\left (11-49\eta+52\eta^2\right
)v^6-\frac{35}{16}\eta(1-5\eta+5\eta^2)\dot x^6+\frac{\eta}{4}\left
(75+32\eta-40\eta^2\right )\left (\frac{1}{x}\right )v^4
\right. \nonumber \\
&\left.
+\frac{\eta}{2}\left (158-69\eta-60\eta^2 \right )\left (\frac{1}{x}\right
)\dot x^4-\eta\left (121-16\eta-20\eta^2\right )\left
(\frac{1}{x}\right )v^2 \dot x^2
\right. \nonumber \\
&\left. 
-\frac{3}{8}\eta\left (20-79\eta+60\eta^2\right )v^4\dot
x^2+\frac{15}{8}\eta\left (4-18\eta+17\eta^2\right )v^2\dot x^4\right
\}
\nonumber \\
&+\frac{\bv}{x^2} \dot x\left \{
\left [4+\left
  (\frac{5849}{840}+\frac{123}{32}\pi^2\right )\eta
-25\eta^2-8\eta^3\right ]\left (\frac{1}{x}\right
)^2+\frac{\eta}{8}\left (65-152\eta-48\eta^2\right )v^4
\right. \nonumber \\
&\left. 
+\frac{15}{8}\eta\left (3-8\eta-2\eta^2\right )\dot x^4+\eta\left
(15+27\eta+10\eta^2\right )\left (\frac{1}{x}\right )v^2
\right. \nonumber \\
&\left. 
-\frac{\eta}{6}\left (329+177\eta+108\eta^2\right )\left
(\frac{1}{x}\right )\dot x^2 - \frac{3}{4}\eta\left
(16-37\eta-16\eta^2\right )v^2\dot x^2
\right \} \quad \nonumber
\end{align}
\begin{align}
{\bf a}_{3.5PN}&=-\frac{\bn}{x^2}\left\{
\frac{8}{5}\eta\left (\frac{1}{x}\right )\dot x \left
     [\frac{23}{14}(43+14\eta )\left (\frac{1}{x}\right
       )^2+\frac{3}{28}(61+70\eta)v^4+70\dot x^4
\right.\right. \nonumber \\
&\left.\left.
+\frac{1}{42}\left (519-1267\eta\right )\left (\frac{1}{x}\right
       )v^2+\frac{1}{4}\left (147+188\eta\right )\left (\frac{1}{x}\right
       )\dot x^2-\frac{15}{4}\left (19+2\eta\right )v^2\dot x^2\right ]
\right \}\nonumber \\
&+\frac{\bv}{x^2}\left \{
\frac{8}{5}\eta\left (\frac{1}{x}\right )\left [\frac{1}{42}\left
  (1325+546\eta\right )\left (\frac{1}{x}\right )^2+\frac{1}{28}\left
  (313+42\eta \right )v^4+75\dot x^4
\right.\right.\nonumber \\
&\left.\left.
-\frac{1}{42}\left (205+777\eta\right )\left (\frac{1}{x}\right
)v^2+\frac{1}{12}\left (205+424\eta \right )\left (\frac{1}{x}\right )\dot
x^2-\frac{3}{4}\left (113+2\eta\right )v^2\dot x^2\right ]
\right \} \quad .
\end{align}
using the 3PN and half-order terms from Ref.\
\cite{Pati:2002ux}.

We also absorb an $M^2$ into the spins so the physical spin in the
Lagrangian system is denoted
${\bf \mathscr S}_i={\avec}_i(m_i^2/M^2)$ where ${\avec}_i$ is the
dimensionless amplitude $0\le |{\avec}_i |\le 1$.  As we will see in \S \ref{sec:ham},
this is a different normalization
from that for spins in the Hamiltonian case, where spins are expressed as
${\bf S}_i={\avec}_i(m_i^2/\mu M)$.
The spin-orbit contribution to the acceleration is
\begin{equation}
\nonumber
{\bf a}_{PN-SO}=\frac{1}{x^3}\left \{
6\frac{\bn}{x}\blnt\cdot(\bsL+\bxi)-\bv\times(4\bsL+3\bxi)+3\dot
x\bn\times (2\bsL+\bxi)
\right \} \nonumber 
\end{equation}
with $\bxi\equiv (m_2/m_1)\bsoneL+(m_1/m_2)\bstwoL$
\cite{Will:2005sn}.
It is customary to define a {\it reduced} Newtonian orbital angular momentum,
\begin{equation}
{\bf \tilde{\bf L}}_N=\bx\times \bv \quad .
\end{equation}
The spin-orbit dissipation term with the covariant spin supplementary condition is
(see the appendix of Ref.\ \cite{Will:2005sn}),
\begin{align}
{\bf a}_{3.5PN-SO}&=-\frac{\eta}{5x^4}\left\{
\frac{\dot x\bn}{x}\left [\left (120v^2+280\dot
  x^2+453\frac{1}{x}\right )\blnt\cdot \bsL
\right.\right. \nonumber \\
&\left.\left.
+\left (120v^2+280\dot
  x^2+458\frac{1}{x}\right )\blnt\cdot \bxi
\right ]\right.
\nonumber \\
&\left. +\frac{\bv}{x}\left [
\left (87v^2-675\dot x^2-\frac{901}{3}\frac{1}{x}\right )\blnt\cdot
\bsL+4\left (18v^2-150\dot x^2-66\frac{1}{x}\right )\blnt\cdot \bxi\right
]
\right. \nonumber \\
&\left. -\frac{2}{3}\dot x \bv \times \bsL\left (48v^2+15\dot
x^2+364\frac{1}{x}\right ) +\frac{1}{3}\dot x \bv \times \bxi \left
(291v^2-705\dot x^2-772\frac{1}{x}\right )
\right. \nonumber \\
&\left. +\frac{1}{2}\bn\times \bsL\left (31v^4-260v^2\dot x^2+245\dot
x^4-\frac{689}{3}v^2\frac{1}{x}+537\dot
x^2\frac{1}{x}+\frac{4}{3}\frac{1}{x^2}\right )
\right. \nonumber \\
&\left.+\frac{1}{2}\bn\times\bxi\left (115v^4-1130v^2\dot x^2+1295\dot
x^4-\frac{869}{3}v^2\frac{1}{x}+849\dot x^2\frac{1}{x}+\frac{44}{3}\frac{1}{x^2}\right )
\right \} \quad .
\end{align}

And the
spin-spin contribution is
\begin{equation}
{\bf a}_{PN-SS}=-\frac{3}{\mu x^4}\left [\bn
  (\bsoneL\cdot\bstwoL)+\bsoneL(\bn\cdot\bstwoL)+\bstwoL(\bn\cdot
  \bsoneL)-5\bn(\bn\cdot \bsoneL)(\bn\cdot\bstwoL)\right ]
\end{equation}
See also \cite{kidder1995}.
The 3.5PN order effects of Spin-Spin coupling from Ref.\
\cite{Wang:2007ntb} are
\begin{align}
{\bf a}_{3.5PN-SS}=&\frac{1}{x^5}\left \{
\bn\left [\left (287\dot x^2-99v^2+\frac{541}{5}\frac{1}{x}\right
  )\dot x(\bsoneL\cdot\bstwoL)-\left (2646\dot
  x^2-714v^2+\frac{1961}5{}\frac{1}{x}\right )\dot x(\bn\cdot
  \bsoneL)(\bn\cdot \bstwoL)
\right.\right.\nonumber \\
&\left.\left. 
+\left (1029\dot
  x^2-123v^2+\frac{629}{10}\frac{1}{x}\right )\left
  ((\bn\cdot\bsoneL)(\bv\cdot
  \bstwoL)+(\bn\cdot\bstwoL)(\bv\cdot\bsoneL)\right ) -336\dot x(\bv
  \cdot \bsoneL)(\bv\cdot \bstwoL)\right ]
\right. \nonumber \\
&\left. 
+\bv \left [\left (\frac{171}{5}v^2-195\dot
  x^2-67\frac{1}{x}\right )(\bsoneL\cdot \bstwoL)-\left (174v^2-1386\dot
  x^2-\frac{1038}{5}\frac{1}{x}\right
  )(\bn\cdot\bsoneL)(\bn\cdot\bstwoL)
\right.\right.\nonumber \\
&\left.\left.
-438\dot x\left
((\bn\cdot\bsoneL)(\bv\cdot\bstwoL)+(\bn\cdot\bstwoL)(\bv\cdot\bsoneL)\right
)+96(\bv\cdot\bsoneL)(\bv\cdot\bstwoL)\right ]
\right.\nonumber \\
&\left.
+\left(\frac{27}{10}v^2-\frac{75}{2}\dot
x^2-\frac{509}{30}\frac{1}{x}\right )\left ((\bv\cdot
\bstwoL)\bsoneL+(\bv\cdot\bsoneL)\bstwoL\right )
\right.\nonumber \\
&\left.
+\left (\frac{15}{2}v^2+\frac{77}{2}\dot
x^2+\frac{199}{10}\frac{1}{x}\right )\dot x\left ((\bn\cdot
\bstwoL)\bsoneL+(\bn\cdot\bsoneL)\bstwoL\right )
\right \} \quad .
\label{3.5PNSS}
\end{align}
The spin supplementary condition has no effect on spin-spin terms up to this order.
We have not yet included the quadrupole-monopole contribution.
Finally,
the spins precess according to
\begin{eqnarray}
\dot \bsoneL&=&\eta \frac{(\bx\times\bv)\times \bsL_1}{x^3}\left
(2+\frac{3m_2}{2m_1}\right ) \nonumber \\
\dot \bstwoL&=&\eta \frac{(\bx\times\bv)\times \bsL_2}{x^3}\left
(2+\frac{3m_1}{2m_2}\right ) \ \ .
\label{kwwso}
\end{eqnarray}
We could also add to the right hand side of (\ref{kwwso}) the spin-spin
terms:
\begin{equation}
(\dot\bsoneL)_{PN-SS}=-\frac{1}{x^3}\left
(\bstwoL-3(\bn\cdot\bstwoL)\bn\right )\times\bsoneL
\end{equation}
\cite{kidder1995}
and
\begin{equation}
(\dot\bsoneL)_{3.5PN-SS}=\frac{\eta}{x^5}\left
  (\frac{2}{3}(\bv\cdot\bstwoL
)+30\dot x(\bn\cdot\bstwoL)\right
  )(\bn\times\bsoneL)
\label{willss}
\end{equation}
\cite{Wang:2007ntb}.
Equations (\ref{keom1})-(\ref{willss}) constitute the Lagrangian
dynamical system. 
These equations are complete through 3.5PN order except for the spin
contributions, for which additional terms have been calculated (see
for instance Ref.\ \cite{2009PhRvD..79j4023A,faye2006}). Since spin effects are already small at
the order we include here, we have not pursued inclusion of the higher-
order spin terms.

Notice these spins are not reduced by $\mu$ and a definition of
dimensionful angular momentum
will have a $\mu$ in it: ${\bf J}=\mu\blnt+...+\bsoneL+\bstwoL$.

Next we re-express the
dissipation terms in the language of the Hamiltonian system.

\section{Hamiltonian Equations of Motion: Including Dissipation}
\label{sec:ham}

The Hamiltonian PN-formulation is by definition conservative and
therefore does not incorporate dissipation \cite{{schaefer1985},{damour1981},{damour1988},{jaranowski1998},{damourpn2001},{damourpn2000:2}}. 
We can, however, take the
half-order acceleration terms from the Lagrangian coordinates and
simply convert them to the coordinates appropriate for the Hamiltonian.
We begin by first laying out the whole-order terms in the usual
Hamiltonian system \cite{{schaefer1985},{damour1981},{damour1988},{jaranowski1998},{damourpn2001},{damourpn2000:2}}. 

In a Hamiltonian formulation, the equations of motion are derived from
\begin{equation}
\dot {\br}=\frac{\partial H}{\partial {\bp}} \quad ,\quad
\dot{\bp}=-\frac{\partial{H}}{\partial {\br}}\quad .
\label{hameoms}
\end{equation}
As is standard convention, we work in dimensionless coordinates: 
the dimensionless coordinate vector, ${\bf r}$, is measured in units of
total mass, $M=m_1+m_2$, for a pair with black
hole masses $m_1$ and $m_2$.
The canonical
momentum,
 ${\bf p}$, is measured in units of the reduced mass,
$\mu=m_1m_2/M$. The dimensionless combination
$\eta =\mu/M$ will again prove useful. We write vector quantities
in bold. The coordinate $r$ is to be understood as the magnitude 
$r=\sqrt{\br\cdot \br}$. Unit vectors such as $\bn = \br/r$ will additionally
carry a hat.
Finally, we have used the 
dimensionless reduced Hamiltonian $H={\cal H}/\mu$ in Eqs.\ (\ref{hameoms}),
where ${\cal H}$ is the physical Hamiltonian, to 3PN
order plus spin-orbit terms 
\cite{{schaefer1985},{damour1981},{damour1988},{jaranowski1998},{damourpn2001},{damourpn2000:2}}. 
$H$ can be expanded as
\begin{equation}
H= H_N+H_{1PN}+H_{2PN}+H_{3PN}+H_{SO}+H_{SS} \quad ,
\label{ham1}
\end{equation}
where
\begin{widetext}
\begin{align}
\label{Eq:HPN}
H_N&=\frac{{\bp}^2}{2}-\frac{1}{r} &  & \\
H_{1PN}&=\frac{1}{8}\left (3\eta-1\right ) \left ({\bp}^2\right )^2
-\frac{1}{2}\left [\left (3+\eta\right ){\bp}^2+\eta({\bn} \cdot 
{\bf  p})^2\right ] \frac{1}{r}+\frac{1}{2r^2} &\nonumber \\
H_{2PN}&= \frac{1}{16} \left (1-5\eta+5\eta^2\right ) \left ({\bp}^2\right )^3
+\frac{1}{8}\left [\left (5-20\eta-3\eta^2\right)\left ({\bp}^2\right )^2 \right.&\nonumber \\
&\left.- 2\eta^2({\bn} \cdot {\bp})^2{\bp}^2
-3\eta^2({\bn} \cdot {\bp})^4\right ] \frac{1}{r} &\nonumber \\
& +\frac{1}{2}\left [\left (5+8\eta\right ){\bp}^2
+3\eta({\bn} \cdot {\bp})^2\right ] \frac{1}{r^2}
-\frac{1}{4}\left( 1+3\eta \right)\frac{1}{r^3} &
\nonumber \\
H_{3PN}&=\frac{1}{128}\left (-5+35\eta-70\eta^2+35\eta^3\right )
\left ({\bp}^2\right )^4
+\frac{1}{16}\left [\left (-7+42\eta-53\eta^2-5\eta^3\right )\left
  ({\bp}^2\right )^3\right. &\nonumber \\
& \left. +(2-3\eta)\eta^2({\bn} \cdot {\bp})^2({\bp}^2)^2
+3(1-\eta)\eta^2({\bn} \cdot {\bp})^4{\bp}^2-5\eta^3({\bn} \cdot
{\bp})^6\right ]\frac{1}{r}&\nonumber \\
&+\left [\frac{1}{16}(-27+136\eta+109\eta^2)({\bp}^2)^2
+\frac{1}{16}(17+30\eta)\eta({\bn} \cdot {\bp})^2{\bp}^2 +
\frac{1}{12}(5+43\eta)\eta({\bn} \cdot {\bp})^4\right ]
\frac{1}{r^2}&
\nonumber \\
&+\left \{ \frac{1}{192}\left [-600+\left
  (3\pi^2- 1340 \right  )
\eta-552\eta^2\right ]{\bp}^2 
-\frac{1}{64}\left(340+3\pi^2+112\eta\right )\eta
({\bn} \cdot {\bp})^2\right \} \frac{1}{r^3}&
\nonumber \\
&+\frac{1}{96}\left [ 12+\left (872-63\pi^2\right )
  \eta\right ]\frac{1}{r^4}\quad ,&
\nonumber \\
H_{SO}&=\frac{\bl\cdot \bs}{r^3} \quad .
\label{Hterms}
\end{align}
\end{widetext}
and adding spin-spin,
\begin{align}
H_{SS}=\frac{\mu}{r^3}\left
[ \right. & \left. 3(\bsone\cdot\bn)(\bstwo\cdot\bn)-\bsone\cdot\bstwo +
\right. \\
&\left. \frac{m_2}{2m_1}\left (3(\bsone\cdot\bn)(\bsone\cdot\bn)-\bsone\cdot\bsone
\right )
+\frac{m_1}{2m_2}\left(3(\bstwo\cdot\bn)(\bstwo\cdot\bn)-\bstwo\cdot\bstwo
\right )
\right ]
\quad .
\nonumber 
\end{align}
Notice that $H_{SS}$ includes $S_1^2$ and $S_2^2$ terms.
For two spinning black holes $\bs$ is\footnote{The definitions for
  $\bs$ can vary in the literature up to an overall constant although
  the reduced $H_{SO}$ must be the same for all prescriptions.}
\begin{equation}
\bs=\delta_1\bsone +\delta_2\bstwo
\label{Eq:Seff}
\end{equation}
where the dimensionless reduced spins are defined as
\begin{equation}
\bsone={\bf {\it a}_1} (m_1^2/\mu M)\, , \quad \quad
\bstwo={\bf {\it a}_2} (m_2^2/\mu M)\, .
\end{equation} 
and
\begin{equation}
\delta_1\equiv\left (2+\frac{3m_2}{2m_1}\right )\eta \, , \quad\quad
\delta_2\equiv\left (2+\frac{3m_1}{2m_2}\right )\eta \, .
\end{equation}
The dimensionless spin amplitudes are confined to the range
$0\le a_{1,2}\le 1$.
The
spins 
precess
according to
\begin{equation}
\frac{d\bsone}{dt}=\frac{\partial H}{\partial \bsone}\times \bsone
\end{equation}
or
\begin{align}
{\bf \dot{{\bf S}}_1}  &= 
\delta_1\frac{\bl\times \bsone}{r^3}
+\frac{\mu}{r^3}\left [3\bn
\left (\left (\bstwo+\frac{m_2}{m_1}\bsone\right )\cdot\bn\right )-\bstwo
\right ]\times \bsone
\nonumber \\
{\bf \dot{{\bf S}}_2}  &= 
\delta_2\frac{\bl\times \bstwo}{r^3} 
+\frac{\mu}{r^3}\left [3\bn
\left (\left (\bsone+\frac{m_1}{m_2}\bstwo\right )\cdot\bn\right )-\bsone
\right ]\times \bstwo
\quad .
\label{ds2}
\end{align}
The orbital angular momentum precesses according to 
\begin{align}
\dot \bl &=\dot \br \times \bp +\br \times \dot \bp \\
&=-\frac{\bl \times \bs}{r^3}-\frac{\mu}{r^3}\left [ 3\bn\left (\left
  (\bstwo +\frac{m_2}{m_1}\bsone\right )\cdot \bn\right )\right
]\times \bsone
-\frac{\mu}{r^3}\left [ 3\bn\left (\left
  (\bsone +\frac{m_1}{m_2}\bstwo\right )\cdot \bn\right )\right
]\times \bstwo\nonumber 
\end{align}
Adding these together, it follows that the total angular momentum
$\bj=\bl+\bsone+\bstwo$
is
conserved 
in the absence of dissipation so that the orbital angular momentum
precesses according to 
\begin{equation}
\dot \bl=-\dot \bsone-\dot \bstwo \quad .
\end{equation}

Now, we are ready to convert the radiation-reaction terms of
appendix \ref{sec:lag} into Hamiltonian variables. To do so, we need
to relate the Hamiltonian variables $(\br,\bp)$ to the Lagrangian
variables $(\bx,\bv)$. 
To convert the 2.5PN radiation-reaction term,
we only need the 1PN-order coordinate conversion to catch all corrections up
to 3.5PN. To convert the 3.5PN radiation-reaction term, we only need
the zeroth-order PN coordinate conversion.
It has been shown that to 1PN order \cite{PhysRevD.63.044021}
\begin{equation}
\br = \bx \quad ,
\end{equation}
so that 
\begin{equation}
\dot \br =\dot \bx =\bv \quad .
\end{equation}
What we really want are the harmonic variables of the Lagrangian formulation
$(\bx,\bv)$ in terms of canonical Hamiltonian variables $(\br,\bp)$
to 1PN order. We have $\bx(\br,\bp)$ above. To find $\bv(\br,\bp)$,
we use Hamilton's equations:
\begin{align}
\dot{\br}&=A\bp +B\bn + \frac{\bs\times \br}{r^3} \nonumber \\
\dot{\bp}&=C\bp +D\bn +  \frac{\bs\times \bp}{r^3} 
+\frac{3\bn}{r} H_{SO} -\frac{\partial H_{SS}}{\partial \br}
\quad ,
\label{eomshort2}
\end{align}
where $A,B,C,D$ are 
\begin{align}
A& \equiv 2\left.\frac{\partial H_{PN}}{\partial \bp^2} \right |_{r,(\bn\cdot\bp)}\\
B& \equiv \left.\frac{\partial H_{PN}}{\partial (\bn\cdot \bp)}\right |_{r,\bp} \nonumber \\
C&\equiv  -\frac{1}{r}\left.\frac{\partial H_{PN}}{\partial (\bn\cdot \bp)}\right |_{r,\bp} =-\frac{B}{r}\nonumber \\
D&\equiv -\left.\frac{\partial H_{PN}}{\partial r}\right |_{\bp,(\bn\cdot\bp)}+\left.\frac{\partial H_{PN}}{\partial
  (\bn\cdot\bp)}\right |_{r,\bp}\frac{(\bn \cdot \bp)}{r}\nonumber \\
&=-\left.\frac{\partial H_{PN}}{\partial r}\right |_{\bp,(\bn\cdot\bp)}
-(\bn \cdot \bp)C \quad .
\label{abcd}
\end{align}
and $H_{PN}=H_N +H_{1PN}+H_{2PN}+H_{3PN}$. Explicit expressions can be found in \cite{Grossman:2008yk}, but
we will only need $A$ and $B$ here
to 1PN order to calculate $\bv(\br,\bp)$ to 1PN order and ultimately
find the radiative contributions to the accelerations. Taking the appropriate
derivatives of the Hamiltonian to 1PN order, we have
\begin{align}
A_{\le 1PN} &=1+
\frac{1}{2}\left (3\eta -1 \right ) \bp^{2}-\left (  3+\eta
\right)\frac{1}{r}
\nonumber \\
B_{\le 1PN} &=
- \eta \left (\bn\cdot \bp  \right)\frac{1}{r} \quad .
\end{align}
This gives us $\bv$ in terms of $(\br,\bp)$:
\begin{equation}
\bv = \left (1+
\frac{1}{2}\left (3\eta -1 \right ) \bp^{2}-\left (  3+\eta
\right)\frac{1}{r}\right )\bp - \eta \left (\bn\cdot \bp
\right)\frac{1}{r} \bn+\frac{\bs\times \br}{r^3}
\label{vconvert}
\end{equation}
and
\begin{align}
v^2 &= \left (1+
\left (3\eta -1 \right ) \bp^{2}-2\left (  3+\eta
\right)\frac{1}{r}\right )\bp^2 - 2\eta \left (\bn\cdot \bp
\right)^2\frac{1}{r} +2\frac{\bp\cdot\left (\bs\times \br \right )}{r^3} \nonumber \\
&= \left (1+
\left (3\eta -1 \right ) \bp^{2}-2\left (  3+\eta
\right)\frac{1}{r}\right )\bp^2 - 2\eta \left (\bn\cdot \bp
\right)^2\frac{1}{r} +2\frac{\bs\cdot\bl }{r^3}
 \quad .
\end{align}
We can now re-write ${\bf a}_{2.5PN}$ of
Eq.\ (\ref{keom2}) in terms
of ADM variables as 
\begin{align}
&{\bf a}_{2.5PN\rightarrow ADM}=
\frac{8}{5}\left (\frac{\eta}{r}\right)\times \left \{
\right. \label{a2.5pnadmrdot} \\
&\left. \frac{\bn}{r^2}
\left [\dot r\left(\frac{17}{3r}+3\bp^{2}
+3\left (3\eta -1 \right ) \bp^{4}-6\left (  3+\eta
\right)\frac{\bp^2}{r} - 6\eta \frac{\left (\bn\cdot \bp\right)^2}{r} 
+6 \frac{\bs\cdot\bl }{r^3} \right )
+ \eta \left (\bn\cdot \bp
\right)
\left (\frac{3}{r^2}+\frac{\bp^2}{r}\right )
\right ]
\right.
\nonumber \\
&\left. -\frac{\bp}{r^2}
\left [\frac{3}{r}+\bp^2+\frac{3}{2}\left (3\eta -1 \right ) \bp^{4}
+\half\left (  3\eta-21\right)\frac{\bp^2}{r} 
-3(3+\eta)\frac{1}{r^2}
- 2\eta \frac{\left (\bn\cdot \bp\right)^2}{r} 
+2 \frac{\bs\cdot\bl }{r^3} 
\right ] \right. \nonumber \\
&\left. -\frac{\bs\times \br}{r^5}\left (\frac{3}{r}+\bp^2\right )
\right \} \quad .
\nonumber
\end{align}
We can also replace $\dot r$ with 
\begin{align}
\dot r=\bn\cdot \bv &=
 \left (1+
\frac{1}{2}\left (3\eta -1 \right ) \bp^{2}-\left (  3+\eta
\right)\frac{1}{r}\right )\bn\cdot\bp - \eta \left (\bn\cdot \bp
\right)\frac{1}{r} \nonumber \\
&=
 \left (1+
\frac{1}{2}\left (3\eta -1 \right ) \bp^{2}-\left (  3+2\eta
\right)\frac{1}{r}\right )\bn\cdot\bp 
\label{rdt}
\end{align}
To get
\begin{align}
&{\bf a}_{2.5PN\rightarrow ADM}=
\frac{8}{5}\left (\frac{\eta}{r}\right)\times \left \{
\right. \label{a2.5pnadm} \\
&\left. \frac{\bn}{r^2}
\left [(\bn\cdot\bp)\left(\frac{17}{3r}+3\bp^{2}
+\frac{9}{2}\left (3\eta -1 \right ) \bp^{4}-\half\left (  5\eta +\frac{179}{3}
\right)\frac{\bp^2}{r} -\frac{1}{3}\left (25\eta+51\right )\frac{1}{r^2}
- 6\eta \frac{\left (\bn\cdot \bp\right)^2}{r} 
+6 \frac{\bs\cdot\bl }{r^3} 
\right )
\right ]
\right.
\nonumber \\
&\left. -\frac{\bp}{r^2}
\left [\frac{3}{r}+\bp^2+\frac{3}{2}\left (3\eta -1 \right ) \bp^{4}
+\half\left (  3\eta-21\right)\frac{\bp^2}{r} 
-3(3+\eta)\frac{1}{r^2}
- 2\eta \frac{\left (\bn\cdot \bp\right)^2}{r} 
+2 \frac{\bs\cdot\bl }{r^3} 
\right ] \right. \nonumber \\
&\left. -\frac{\bs\times \br}{r^5}\left (\frac{3}{r}+\bp^2\right )
\right \} \quad .
\nonumber
\end{align}
Since we used a coordinate change valid to 1PN order,
Eqs.\ (\ref{a2.5pnadm} and \ref{a2.5pnadmrdot}) includes 3.5PN corrections as well as 2.5PN corrections. 

To rewrite ${\bf a}_{3.5PN\rightarrow ADM}$ we simply take the harmonic expressions
and replace $\bx=\br$ and $\bv=\bp$, noting that $\blnt=\bl$ to lowest order, and
that the spins of Refs.\ \cite{Will:2005sn,Wang:2007ntb} are $\eta$ times those of this
section: $\bsL_i= \eta {\bf S}_i$. 

Now, to incorporate the effects of dissipation, we can add the
appropriately converted accelerations.
Taking the time derivative of Eq.\ (\ref{vconvert}) and rearranging to
solve for $\dot {\bp}$, there are additional corrections of
the form
\begin{align}
\dot \bp & = \dot\bv\left (1-
\frac{1}{2}\left (3\eta -1 \right ) \bp^{2}+\left (  3+\eta
\right)\frac{1}{r}\right ) -  (3\eta
-1)(\bp\cdot\dot\bp)\bp +\eta(\bn\cdot\dot \bp)\frac{1}{r}\bn+...
\end{align}
We then use $\dot \bv={\bf a}_{2.5PN\rightarrow ADM}+{\bf
  a}_{3.5PN\rightarrow ADM}$ and drop all terms higher than 3.5PN to get
\begin{align}
\dot \bp &=...+{\bf a}_{2.5PN\rightarrow ADM}\left (1-
\frac{1}{2}\left (3\eta -1 \right ) \bp^{2}
+\left (  3+\eta\right)\frac{1}{r}\right )
+{\bf a}_{3.5PN\rightarrow  ADM}+
\nonumber \\&
- 
(3\eta-1)(\bp\cdot{\bf a}_{2.5PN})\bp+\eta(\bn\cdot{\bf
  a}_{2.5PN})\frac{1}{r}\bn
\label{moreterms}
\end{align}
where we have only isolated the relevant half-order terms so ``$...$''
represents all the whole-order PN terms. Here ${\bf
  a}_{2.5PN\rightarrow ADM}$ is
 defined in Eq. (\ref{a2.5pnadm}), and we replace
$\dot \bp$ where it occurs on the right hand side of (\ref{moreterms})
with ${\bf a}_{2.5PN}$ evaluated at $\bx=\br$,
 $\bv=\bp$.
The term ${\bf
  a}_{3.5PN\rightarrow ADM}$ is also evaluated at $\bx=\br$,
 $\bv=\bp$, and, again, the spins of the previous section are related to these
spins through $\bsL_i= \eta {\bf S}_i$, 
yielding
\begin{align}
{} &{\bf \tilde a}_{2.5PN\rightarrow ADM}\equiv
 {\bf a}_{2.5PN\rightarrow ADM}\left (1-\frac{1}{2}\left (3\eta -1 \right ) \bp^{2} 
+\left (  3+\eta \right)\frac{1}{r}\right )=\nonumber \\
&\frac{8}{5}\left (\frac{\eta}{r}\right)\times \left \{
\frac{\bn}{r^2}
\left [\bn\cdot\bp\left( \frac{17}{3r}+3\bp^{2}
+3\left (3\eta -1 \right ) \bp^{4}-2\left (4 \eta+9
\right)\frac{\bp^2}{r} -\frac{8}{3}\frac{\eta}{r^2}- 6\eta \frac{\left (\bn\cdot \bp
\right)^2}{r} +6 \frac{\bs\cdot\bl }{r^3} \right )
\right ]
\right. \nonumber \\ 
&\left. 
-\frac{\bp}{r^2}
\left [\frac{3}{r}+\bp^2+\left (3\eta -1 \right ) \bp^{4}
-2\left (  \eta+3\right)\frac{\bp^2}{r} 
-2\eta \frac{\left (\bn\cdot \bp\right)^2}{r} 
+2 \frac{\bs\cdot\bl }{r^3} 
\right ]
\right.\nonumber \\ 
&\left.  -\frac{\bs\times \br}{r^5}\left (\frac{3}{r}+\bp^2\right )
\right \} \quad ,
\end{align}

Finally, our equations of motion become
\begin{align}
\dot{\br}&=A\bp +B\bn + \frac{\bs\times \br}{r^3} \nonumber \\
\dot{\bp}&=C\bp +D\bn +  \frac{\bs\times \bp}{r^3} 
+\frac{3\bn}{r} H_{SO} -\frac{\partial H_{SS}}{\partial \br}
+
{\bf \tilde a}_{2.5PN\rightarrow ADM}+{\bf a}_{3.5PN\rightarrow ADM} &
\nonumber \\
&
+\eta(\bn\cdot{\bf a}_{2.5PN})\frac{1}{r}\bn
-(3\eta-1)(\bp\cdot{\bf a}_{2.5PN})\bp
\label{eomsdiss}
\end{align}

We can regroup the accelerations to 2.5 order and
3.5 orders in ADM variables:
\begin{align}
\dot{\br}&=A\bp +B\bn + \frac{\bs\times \br}{r^3} \nonumber \\
\dot{\bp}&=C\bp +D\bn +  \frac{\bs\times \bp}{r^3} 
+\frac{3\bn}{r} H_{SO} -\frac{\partial H_{SS}}{\partial \br}
+{\bf a}^{(2.5)}_{ADM}+{\bf a}^{(3.5)}_{ADM}
\label{eomsdissADM}
\end{align}
with 
\begin{align}
{\bf a}^{(2.5)}_{ADM}=
&\frac{8}{5}\left (\frac{\eta}{r}\right)\times \left \{
\frac{\bn}{r^2}
\left [\bn\cdot\bp\left( \frac{17}{3r}+3\bp^{2}
 \right )
\right ]
-\frac{\bp}{r^2}
\left [\frac{3}{r}+\bp^2
\right ]
\right \} \quad ,
\end{align}
and using
\begin{align}
\eta(\bn\cdot{\bf
  a}_{2.5PN})\frac{1}{r}\bn &=
\left (\frac{8\eta}{5r}\right
)\frac{\bn}{r^2} (\bn\cdot\bp)\left (\frac{8\eta}{3r^2}+2\eta\frac{\bp^2}{r}\right )
\nonumber \\
- 
(3\eta-1)(\bp\cdot{\bf a}_{2.5PN})\bp &=\left (\frac{8\eta}{5r}\right
) \left (- \frac{\bp}{r^2}\right )  (3\eta-1)\left [(\bn\cdot\bp)^2\left
  (\frac{17}{3r}+3\bp^2\right )-\bp^2\left (\frac{3}{r}+\bp^2\right
  )\right ]
\end{align}
and the 3.5PN piece in $\tilde a_{2.5PN\rightarrow ADM}$
\begin{align}
...
&\frac{8}{5}\left (\frac{\eta}{r}\right)\times \left \{
\frac{\bn}{r^2}
\left [\bn\cdot\bp\left( 
3\left (3\eta -1 \right ) \bp^{4}-2\left (4 \eta+9
\right)\frac{\bp^2}{r} -\frac{8}{3}\frac{\eta}{r^2}- 6\eta \frac{\left (\bn\cdot \bp
\right)^2}{r} +6 \frac{\bs\cdot\bl }{r^3} \right )
\right ]
\right. \nonumber \\ 
&\left. 
-\frac{\bp}{r^2}
\left [\left (3\eta -1 \right ) \bp^{4}
-2\left (  \eta+3\right)\frac{\bp^2}{r} 
-2\eta \frac{\left (\bn\cdot \bp\right)^2}{r} 
+2 \frac{\bs\cdot\bl }{r^3} 
\right ]
\right.\nonumber \\ 
&\left.  -\frac{\bs\times \br}{r^5}\left (\frac{3}{r}+\bp^2\right )
\right \} \quad ,
\end{align}
and adding all of these 3.5PN pieces to
\begin{align}
{\bf a}_{3.5PN\rightarrow ADM}
&=\frac{8}{5}\eta\left (\frac{1}{r}\right )\times\left \{
\right. \nonumber \\
&\left.-\frac{\bn}{r^2}
(\bn\cdot\bp) \left
     [\frac{23}{14}(43+14\eta )\left (\frac{1}{r}\right
       )^2+\frac{3}{28}(61+70\eta)\bp^4+70(\bn\cdot\bp)^4
\right.\right. \nonumber \\
&\left.\left.
+\frac{1}{42}\left (519-1267\eta\right )\left (\frac{1}{r}\right
       )\bp^2+\frac{1}{4}\left (147+188\eta\right )\left (\frac{1}{r}\right
       )(\bn\cdot\bp)^2-\frac{15}{4}\left (19+2\eta\right )\bp^2(\bn\cdot\bp)^2
\right ]\right.\nonumber \\
&\left.+\frac{\bp}{r^2}\left [
\frac{1}{42}\left
  (1325+546\eta\right )\left (\frac{1}{r}\right )^2+\frac{1}{28}\left
  (313+42\eta \right )\bp^4+75(\bn\cdot\bp)^4
\right.\right.\nonumber \\
&\left.\left.
-\frac{1}{42}\left (205+777\eta\right )\left (\frac{1}{r}\right
)\bp^2+\frac{1}{12}\left (205+424\eta \right )\left (\frac{1}{r}\right )\dot
r^2-\frac{3}{4}\left (113+2\eta\right )\bp^2(\bn\cdot\bp)^2\right ]
\right \} \quad .\nonumber
\end{align}
gives
\begin{align}
&{\bf a}^{(3.5)}_{ADM}
=\frac{8}{5}\eta\left (\frac{1}{r}\right )\times\left \{
\right. \nonumber \\
&\left.-\frac{\bn}{r^2}
(\bn\cdot\bp) \left
     [\frac{1}{14}\left (989+322\eta \right )\left (\frac{1}{r}\right
       )^2+\frac{3}{28}\left (89-14\eta\right )\bp^4+70(\bn\cdot\bp)^4
\right.\right. \nonumber \\
&\left.\left.
+\frac{1}{42}\left (1275-1015\eta\right )\left (\frac{1}{r}\right
       )\bp^2+\frac{1}{4}\left (147+212\eta\right )\left (\frac{1}{r}\right
       )(\bn\cdot\bp)^2-\frac{15}{4}\left (19+2\eta\right )\bp^2(\bn\cdot\bp)^2
-6 \frac{\bs\cdot\bl }{r^3} \right ]\right.\nonumber \\
&\left.+\frac{\bp}{r^2}\left [
\frac{1}{42}\left
  (1325+546\eta\right )\left (\frac{1}{r}\right )^2+\frac{1}{28}\left
  (313+42\eta \right
)\bp^4+75(\bn\cdot\bp)^4+\frac{1}{12}\left(273+244\eta\right) \frac{(\bn\cdot\bp)^2}{r}
\right.\right.\nonumber \\
&\left.\left.
-\frac{1}{42}\left (79+315\eta\right )\left (\frac{1}{r}\right
)\bp^2
-\frac{1}{4}\left (327+42\eta\right )\bp^2(\bn\cdot\bp)^2
-2\frac{\bs\cdot\bl }{r^3} 
\right ]\right. \nonumber \\
&\left.  -\frac{\bs\times \br}{r^5}\left (\frac{3}{r}+\bp^2\right )
\right \} \quad .\nonumber
\end{align}

Then there are the spin pieces:
\begin{align}
&{\bf a}^{(3.5PN-SO)}_{ADM}=-\frac{\eta^2}{5r^4}\left\{
\frac{(\bn\cdot\bp)\bn}{r}\left [\left
  (120\bp^2+280(\bn\cdot\bp)^2+453\frac{1}{r}\right ) \bl\cdot \bstot
\right.\right. \nonumber \\
&\left.\left.
+\left (120\bp^2+280(\bn\cdot\bp)^2+458\frac{1}{r}\right )\bl\cdot \bri
\right ]\right.
\nonumber \\
&\left. +\frac{\bp}{r}\left [
\left (87\bp^2-675(\bn\cdot\bp)^2-\frac{901}{3}\frac{1}{r}\right )\bl\cdot
\bstot+4\left (18\bp^2-150(\bn\cdot\bp)^2-66\frac{1}{r}\right )\bl\cdot \bri\right
]
\right. \nonumber \\
&\left. -\frac{2}{3}(\bn\cdot\bp) \bp\times \bstot\left (48\bp^2+15(\bn\cdot\bp)^2+364\frac{1}{r}\right ) +\frac{1}{3}(\bn\cdot\bp) \bp \times \bri \left
(291\bp^2-705(\bn\cdot\bp)^2-772\frac{1}{r}\right )
\right. \nonumber \\
&\left. +\frac{1}{2}\bn\times \bstot\left (31\bp^4-260\bp^2(\bn\cdot\bp)^2+245(\bn\cdot\bp)^4-\frac{689}{3}\bp^2\frac{1}{r}+537(\bn\cdot\bp)^2\frac{1}{r}+\frac{4}{3}\frac{1}{r^2}\right )
\right. \nonumber \\
&\left.+\frac{1}{2}\bn\times\bri\left (115\bp^4-1130\bp^2(\bn\cdot\bp)^2+1295(\bn\cdot\bp)^4-\frac{869}{3}\bp^2\frac{1}{r}+849(\bn\cdot\bp)^2\frac{1}{r}+\frac{44}{3}\frac{1}{r^2}\right )
\right \} \quad .
\end{align}
and
\begin{align}
&{\bf a}^{(3.5PN-SS)}_{ADM}=\frac{\eta^2}{r^5}\left \{
\bn\left [\left (287(\bn\cdot\bp)^2-99\bp^2+\frac{541}{5}\frac{1}{r}\right
  )(\bn\cdot\bp)(\bsone\cdot\bstwo)\right. \right. \nonumber \\
& \left. \left. -\left (2646(\bn\cdot\bp)^2-714\bp^2+\frac{1961}5{}\frac{1}{r}\right )(\bn\cdot\bp)(\bn\cdot
  \bsone)(\bn\cdot \bstwo)
\right.\right.\nonumber \\
&\left.\left. 
+\left (1029(\bn\cdot\bp)^2-123\bp^2+\frac{629}{10}\frac{1}{r}\right )\left
  ((\bn\cdot\bsone)(\bp\cdot
  \bstwo)+(\bn\cdot\bstwo)(\bp\cdot\bsone)\right ) -336(\bn\cdot\bp)(\bp
  \cdot \bsone)(\bp\cdot \bstwo)\right ]
\right. \nonumber \\
&\left. 
+\bp \left [\left (\frac{171}{5}\bp^2-195(\bn\cdot\bp)^2-67\frac{1}{r}\right )(\bsone\cdot \bstwo)-\left (174\bp^2-1386(\bn\cdot\bp)^2-\frac{1038}{5}\frac{1}{r}\right
  )(\bn\cdot\bsone)(\bn\cdot\bstwo)
\right.\right.\nonumber \\
&\left.\left.
-438(\bn\cdot\bp)\left
((\bn\cdot\bsone)(\bp\cdot\bstwo)+(\bn\cdot\bstwo)(\bp\cdot\bsone)\right
)+96(\bp\cdot\bsone)(\bp\cdot\bstwo)\right ]
\right.\nonumber \\
&\left.
+\left(\frac{27}{10}\bp^2-\frac{75}{2}(\bn\cdot\bp)^2-\frac{509}{30}\frac{1}{r}\right )\left ((\bp\cdot
\bstwo)\bsone+(\bp\cdot\bsone)\bstwo\right )
\right.\nonumber \\
&\left.
+\left (\frac{15}{2}\bp^2+\frac{77}{2}(\bn\cdot\bp)^2+\frac{199}{10}\frac{1}{r}\right )(\bn\cdot\bp)\left ((\bn\cdot
\bstwo)\bsone+(\bn\cdot\bsone)\bstwo\right )
\right \} \quad .
\label{3.5PNSSADM}
\end{align}
with $\bri\equiv (m_2/m_1)\bsone+(m_1/m_2)\bstwo$.

Given the equations of this section and the previous section,  the
equations of motion can be evolved in either set of variables.

In \S \ref{sec:dynamics} we introduced the $(r,\Phi,\Psi)$
coordinates. The equations of motion (\ref{eomsdissADM}) can be projected onto this
basis as was done in Refs.\ \cite{Levin:2008ci,Grossman:2008yk}.

\bibliographystyle{aip}
\bibliography{eomspaper}

\end{document}